\documentclass[12pt]{article}%
\usepackage[nosort]{cite}
\usepackage[usenames, dvipsnames]{xcolor}
\usepackage{graphicx}
\usepackage{multicol}
\usepackage{amsfonts}
\usepackage{amssymb}
\usepackage{amsmath}
\usepackage{heck}
\usepackage{afterpage}
\usepackage{setspace}
\usepackage{verbatim}
\usepackage{color}
\usepackage{longtable}
\usepackage{float}
\usepackage{subcaption}
\usepackage{epsfig}
\usepackage{enumerate}
\usepackage{epstopdf}
\usepackage[enableskew, vcentermath]{youngtab}
\usepackage{adjustbox}
\usepackage{multirow}
\usepackage{tikz}
\usepackage[margin=1in]{geometry}
\usepackage{titletoc}
\usepackage[percent]{overpic}
\usepackage{gensymb}
\usepackage{mathtools}
\usepackage{tikz-cd}%
\providecommand{\U}[1]{\protect\rule{.1in}{.1in}}
\pdfoutput=1
\newsavebox{\mysavebox}

\usetikzlibrary{decorations.markings}

\numberwithin{equation}{section}

\usetikzlibrary{chains}
\allowdisplaybreaks
\tikzset{node distance=2em, ch/.style={circle,draw,on chain,inner sep=2pt},chj/.style={ch,join},every path/.style={shorten >=4pt,shorten <=4pt},line width=1pt,baseline=-1ex}

\newcommand{\ba}{\begin{eqnarray}}
\newcommand{\ea}{\end{eqnarray}}

\newcommand{\be}{\begin{equation}}
\newcommand{\ee}{\end{equation}}

\tikzstyle{startstop} = [rectangle, rounded corners, minimum width=3cm, minimum height=1cm,text centered, draw=black, fill=blue!10]
\tikzstyle{startstop} = [rectangle, rounded corners, minimum width=3cm, minimum height=1cm,text centered, draw=black, fill=blue!10]
\tikzstyle{io} = [trapezium, trapezium left angle=70, trapezium right angle=110, minimum width=3cm, minimum height=1cm, text centered, draw=black, fill=blue!30]
\tikzstyle{process} = [rectangle, minimum width=3cm, minimum height=1cm, text centered, draw=black, fill=orange!30]
\tikzstyle{decision} = [diamond, minimum width=3cm, minimum height=1cm, text centered, draw=black, fill=green!30]
\tikzstyle{arrow} = [thick,->,>=stealth]
\tikzset{->-/.style={decoration={
  markings,
  mark=at position #1 with {\arrow[scale=2.4]{>}}},postaction={decorate}}}
\makeatletter \@addtoreset{equation}{section} \makeatother

\colorlet{darkblue}{blue!70!black}
\colorlet{darkgreen}{green!70!black}

\usepackage[colorlinks=true,urlcolor=darkblue,linktocpage=true,linkcolor=darkblue,citecolor=darkblue]{hyperref}
\hypersetup{
colorlinks=true,
linkcolor=blue,
citecolor=magenta,
}

\begin{document}

\begin{flushright}
QMUL-PH-22-13
\end{flushright}

\date{April 2022}

\title{On the Dynamics of Inference and Learning}

\institution{QM}{\centerline{${}^{1}$Centre for Theoretical Physics, Queen Mary University of London, London E1 4NS, UK}}

\institution{PENN}{\centerline{${}^{2}$Department of Physics and Astronomy, University of Pennsylvania, Philadelphia, PA 19104, USA}}

\institution{PENNMATH}{\centerline{${}^{3}$Department of Mathematics, University of Pennsylvania, Philadelphia, PA 19104, USA}}

\institution{UIUC}{\centerline{${}^{4}$Department of Physics, University of Illinois, Urbana IL 61801, USA}}

\authors{
David S. Berman\worksat{\QM}\footnote{e-mail: {\tt d.s.berman@qmul.ac.uk}},
Jonathan J. Heckman\worksat{\PENN, \PENNMATH}\footnote{e-mail: {\tt jheckman@sas.upenn.edu}} and Marc Klinger\worksat{\UIUC}\footnote{e-mail: {\tt marck3@illinois.edu}}}

\abstract{Statistical Inference is the process of determining a probability distribution over the space of parameters of a model given a data set. As more data becomes available this probability distribution becomes updated via the application of Bayes' theorem. We present a treatment of this Bayesian updating process as a continuous dynamical system. Statistical inference is then governed by a first order differential equation describing a trajectory or flow in the information geometry determined by a parametric family of models. We solve this equation for some simple models and show that when the Cram\'{e}r-Rao bound is saturated the learning rate is governed by a simple $1/T$ power-law, with $T$ a time-like variable denoting the quantity of data. The presence of hidden variables can be incorporated in this setting, leading to an additional driving term in the resulting flow equation. We illustrate this with both analytic and numerical examples based on Gaussians and Gaussian Random Processes and inference of the coupling constant in the 1D Ising model. Finally we compare the qualitative behaviour exhibited by Bayesian flows to the training of various neural networks on benchmarked data sets such as MNIST and CIFAR10 and show how that for networks exhibiting small final losses the simple power-law is also satisfied.}

\maketitle

\setcounter{tocdepth}{2}

\tableofcontents


\newpage

\section{Introduction \label{sec:INTRO}}

It is difficult to overstate the importance of statistical inference. It forms
the bedrock of how one weighs new scientific evidence, and is in some sense
the basis for all of rational thought. Leaving philosophy aside, one can ask
about the mechanics of inference: given new data, how quickly can we expect to
adjust our understanding, and in what sense does this converge to the truth?

Bayes' rule \cite{Bayes:1764vd} provides a concrete way to approach this question. Given events
$A$ and $B$, the conditional probabilities are related as:\footnote{More
symmetrically, one can write $P(A|B)P(B)=P(B|A)P(A).$}%
\begin{equation}
P(A|B) = \frac{P(B|A)P(A)}{P(B)}.
\end{equation}
As new evidence arrives, the posterior $P(A|B)$ can be treated as a
new prior, and Bayes' rule thus provides a concrete way to continue updating
(and hopefully improving) one's initial inference scheme.

In physical applications, one typically imposes a great deal of additional
structure which allows one to weigh the various merits of new evidence. For
example, in the context of quantum field theory, one is often interested in
particle excitations where the structure of locality is built into the
inference scheme. In this context, \textquotedblleft new
evidence\textquotedblright\ amounts to probing shorter distance scales with
the help of a higher energy collider experiment or a more precise measurement
of a coupling constant. In modern terms, this is organized with the help of
the renormalization group \cite{Gell-Mann:1954yli, Kadanoff:1966wm, Wilson:1971bg, Wilson:1971dh, Polchinski:1983gv},
which provides a general way to organize new data as relevant, marginal or irrelevant (in terms of its impact on long distance observables).

More broadly, the issue of identifying relevant features as a function of
scale is an important issue in a range of inference problems. For example, in
machine learning applications, one might wish to classify an image according
to ``dog vs. cat'', and then proceed to breed, and even finer distinguishing
features. In this setting, however, the notion of a single quantity such as
energy / wavenumber to define ``locality'' (as used in quantum field theory)
is far less clear cut. This is also an issue in a wide range of systems with
multiple scales and chaotic dynamics. In these settings it would seem
important to seek out physically anchored organizational principles.

Our aim in this note will be to show that in many circumstances, there is an
emergent notion of scaling which can be traced all the way back to incremental
Bayesian updates. The essential idea is that as an inference scheme converges
towards a best guess, the Bayesian update equation comes to resemble a
diffusion equation. Much as in \cite{Balasubramanian:1996bn}, the appropriate notion of energy
in this context is the Kullback-Leibler divergence \cite{KullbackLeibler} between the model
distribution $m(x)$ and the true distribution $t(x)$:
\begin{equation}
D_{KL}(t||m)=\int dx \, t(x)\mathrm{log}\frac{t(x)}{m(x)}.
\end{equation}
In many applications, the model depends on a set of fitting parameters
$\theta^{i}$, and the problem of inference amounts to performing an update
$m(x|\theta_{k+1})\leftarrow m(x|\theta_{k})$. Indeed, in the infinitesimal
limit where the $\theta_{k}$'s converge to an optimal $\theta_{\ast}^{i}$, we
can expand in the vicinity of this point. The second order expansion in $\theta^i$ of the KL divergence then yields the
Fisher information metric which forms the basis for information geometry \cite{Rao,Amari}. (See  \cite{Bialek:1996kd,Blau:2001gj,Miyamoto:2012qv,Heckman:2013kza,Heckman:2016wte,Heckman:2016jud,Clingman:2015lxa,
Malek:2015hea,Dimov:2020fzi,Erdmenger:2020vmo,Tsuchiya:2021hvg,Fowler:2021oje} for contemporary appearances
of the information metric in statistical inference and its relevance to quantum field theory and string theory.)

The processes of inference through Bayes' theorem then amounts to specifying
the trajectory of a particle in the curved background described by the information geometry.
In Bayesian inference, we treat the parameters $\theta$ as random variables as well, and these are dictated by a
posterior distribution $\pi(\theta , T)$ which updates as a function of data steps $T$. For any observable $O(\theta)$
which depends on these parameters, its appearance in various averages results in
an implicit time dependence and a corresponding flow equation:
\begin{equation}
\frac{\partial}{\partial T} \overline{O}(T) = - \mathrm{Var}(O,D),
\end{equation}
where the line on top indicates an average with respect to $\pi(\theta , T)$, and $\mathrm{Var}(O,D)$ denotes the
variance between the observable and $D$, a KL divergence between the ``true'' distribution and one which depends on $\theta$ at some intermediate stage of the inference scheme. There is a striking formal resemblance between the evolution of the parameters of the model (by taking $O = \theta$), and the evolution of parameters in renormalization group flow. We will explore this analogy further, especially with regards to ``perturbations'' --- i.e. new data --- which can alter the trajectory of a flow.

In favorable circumstances, we can obtain good approximate solutions to this flow equation for a wide class of observables. In fact, for observables where the Cram\'{e}r-Rao bound \cite{Rao} is saturated, we can solve the equation exactly. This then yields a simple $1/T$  power-law scaling. We examine the perturbed flow equation where the Cram\'{e}r-Rao bound is not saturated and solve the equation numerically to give a power-law scaling with powers greater than $-1$. Finally, we examine the case where there are ``hidden variables". These are non-updated parameters in the model and demonstrate an exponential behavior in this case. The interpolation from this simple $1 / T$ to exponential falloff is well-approximated by a power-law decay of the form $1 / T^{1 + \nu}$ with $\nu > 0$.

We illustrate these general considerations with a number of examples. As one of the few cases we can treat analytically,
we illustrate how the flow equations work in the case of inference on data drawn from a Gaussian distribution. This also includes
the important case of a Gaussian Random Process, which is of relevance in the study of (untrained) neural networks in the infinite width limit \cite{Neal}. As a physically motivated numerical example,
we ask how well an observer can learn the coupling constants of the 1D Ising model.
In this setting, ``data'' amounts to sampling from the Boltzmann distribution of possible spin configurations,
and inference corresponds to refining our prior estimates on the value of the nearest neighbor coupling.
We indeed find that the trajectory of the coupling obeys the observable flow equation, and converges to a high level of accuracy.

It is also natural to ask whether we can apply these considerations even when we do not have a generative model for the probability distribution. A classic example of this sort is the ``inference'' performed by a neural network as it is learning. From the Bayesian perspective, neural networks are models that contain a large number of parameters given by the weights and biases of the network and training is a flow on those weights and biases induced by the training data set. See \cite{Mehta:2018dln} for an introduction to neural networks aimed at physicists. Due to the large number of parameters a true quantitative analysis, as we did for the Ising model, is not possible. However, insofar as the neural network is engaging in rational inference, we should expect a flow equation to hold. To test this expectation we study the phenomenology of training a network as a function of the size of the data set. We consider simple dense feedforward networks (FF) and convolutional neural networks (CNNs) trained on the, by now classic, data sets of MNIST, Fashion-MNIST and CIFAR10.\footnote{These data sets are readily available with supporting notes in {\url{https://keras.io/api/data sets/}}}

Quite remarkably, we find that the qualitative $1/T$ power-law behavior is emulated for the MNIST data set where the network final loss is very small and other power-laws occur for more complex data sets where the final loss is higher. The fact that our simple theoretical expectations match on to the rather opaque inference procedure of a neural network lends additional support to the formalism.

The rest of this paper is organized as follows. We begin in section
\ref{sec:BAYES} by reviewing Bayes' rule and then turn to the infinitesimal
version defined by incremental updates and the induced flow for observables. After solving the flow equations exactly and in perturbation theory we then turn to some examples. First, in section
\ref{sec:GAUSS} we present an analytic treatment in the context of inference
for Gaussian data. In section \ref{sec:ISING} we study inference
of coupling constants in the context of the 1D Ising model. In section
\ref{sec:MNIST} we turn to examples of neural network learning various data sets. We present our conclusions and potential avenues for future
investigation in section \ref{sec:CONC}. Appendix \ref{app:PATH} presents a statistical mechanics interpretation of the Bayesian flow equations.

\section{Bayes' Rule as a Dynamical System} \label{sec:BAYES}

In this section we present a physical interpretation of Bayes' rule as a
dynamical system. By working in a limit where we have a large number of events
$N\gg1$ partitioned up into smaller number of events $N_{k}\gg1$ such that
$N_{k}/N \ll 1$, we show that this can be recast as an integro-differential equation.

To begin, suppose we have observed some events $E=\{e_{1},...,e_{N}\}$, as
drawn from some true distribution. In the Bayesian setting, we suppose that we
have some model of the world specified by a posterior distribution conditional
on observed events $f(\theta|e_{1},...,e_{N})=\pi_{\text{post}}(\theta)$,
which depends on \textquotedblleft fitting parameters\textquotedblright%
\ $\theta=\{\theta^{1},...,\theta^{m}\}$ and our density for data conditional
on $\theta$ specified as $f(e_{1},...,e_{N}|\theta)$. In what follows, we
assume that there is a specific value $\theta=\alpha_{\ast}$ for which we
realize the true distribution.\footnote{A word on notation. We have chosen to write the fixed value of a given parameter
as $\alpha_{\ast}$ instead of $\theta_{\ast}$. We do this to emphasize that the $\theta$'s are to be treated as the values of a random variable, with $\alpha_{\ast}$ indicating what a frequentist might refer to as the estimator of this parameter.}
A general comment here is that we are framing our inference problem using Bayesian methods, which means that the parameters
$\theta$ are themselves treated as values drawn from a random distribution.
This is to be contrasted with how we would treat the inference problem as
frequentists, where we would instead attempt to find a best estimate for these
parameters (e.g. the mean and variance of a normal distribution). Rather, the
notion of a Bayesian update means that additional \textquotedblleft
hyperparameters\textquotedblright\ for $\pi_{\mathrm{post}}(\theta)$ are being
updated as a function of increased data. For now, we keep this
dependence on the hyperparameters implicit, but we illustrate later on how
this works in some examples.

Now, assuming the observed events are conditionally independent of $\theta$,
we have:%
\begin{equation}
\pi_{\mathrm{post}}(\theta)=f(\theta|e_{1},...,e_{N})\propto f(e_{1}%
,...,e_{N}|\theta)\times\pi_{\mathrm{prior}}(\theta),
\end{equation}
where the constant of proportionality is fixed by the condition that
$\pi_{\mathrm{post}}(\theta)$ is properly normalized, i.e., we can introduce
another distribution:
\begin{equation}
f(e_{1},...,e_{N})=\int d\theta\,f(e_{1},...,e_{N}|\theta)\pi_{\mathrm{prior}%
}(\theta),
\end{equation}
and write:%
\begin{equation}
\frac{\pi_{\mathrm{post}}(\theta)}{\pi_{\mathrm{prior}}(\theta)}=\frac
{f(e_{1},...,e_{N}|\theta)}{f(e_{1},...,e_{N})}.
\end{equation}

Rather than perform one large update, we could instead consider partitioning
up our events into separate sequences of events, which we can label as
$E(k)=\{e_{1}(k),...,e_{N_{k}}(k)\}$, where now we let $k=1,..,,K$ such that
$N_{1}+...+N_{K}=N$. Introducing the cumulative set of events:
\begin{equation}
S_{k}\equiv E(1)\cup...\cup E(k),
\end{equation}
we can speak of a sequential update, as obtained from incorporating our new
data:
\begin{equation}
\frac{\pi_{k+1}(\theta)}{\pi_{k}(\theta)}=\frac{f(E(k+1)|S_{k},\theta
)}{f(E(k+1)|S_{k})}. \label{recursion}%
\end{equation}

This specifies a recursion relation and thus a discrete dynamical system.
Indeed, writing $\pi_{k}(\theta)=\exp(\ell_{k}(\theta))$ and taking the
logarithm of equation (\ref{recursion}) yields the finite difference equation:%
\begin{equation}
\ell_{k+1}(\theta)-\ell_{k}(\theta)=\log\frac{f(E(k+1)|S_{k},\theta
)}{f(E(k+1)|S_{k})}. \label{finitediff}%
\end{equation}

To proceed further, we now make a few technical assumptions. First of all, we
assume that each draw from the true distribution is independent so that we can
write:%
\begin{equation}
f(e_{1},...,e_{N}|\theta)=f\left(  e_{1}|\theta\right)  ...f\left(
e_{N}|\theta\right)  .
\end{equation}
Further, we assume that draws from the true distribution can always be viewed
as part of the same parametric family as these densities:
\begin{equation}
f(E(k+1)|S_{k})=\underset{e\in E(k+1)}{%
{\displaystyle\prod}
}f(e|\alpha_{k}),
\end{equation}
where $\alpha_k$ is the most likely estimate of $\theta$ given the data $E(k)$.
Then, the finite difference equation of line (\ref{finitediff}) reduces to:%
\begin{equation}
\ell_{k+1}(\theta)-\ell_{k}(\theta)=\underset{e\in E(k+1)}{\sum}\log
\frac{f(e|\theta)}{f(e|\alpha_{k})}\equiv N_{k+1}\left\langle \log
\frac{f(y|\theta)}{f(y|\alpha_{k})}\right\rangle _{E(k+1)},
\label{finitediffagain}%
\end{equation}
in the obvious notation.

We now show that in the limit $N\gg N_{k}\gg1$, Bayesian updating is
well-approximated by an integro-differential flow equation. The large $N_{k}$
limit means that we can approximate the sum on the righthand side of equation
(\ref{finitediffagain}) by an integral:%
\begin{equation}
\ell_{k+1}(\theta)-\ell_{k}(\theta)=N_{k+1}\int dy\text{ }f(y|\alpha_{\ast
})\log\frac{f(y|\theta)}{f(y|\alpha_{k})}+...,
\end{equation}
where the correction terms are subleading in a $1/N_{k+1}$ expansion and we
have switched from referring to events $e_{i}$ by their continuous analogs
$y$. The righthand side can be expressed in terms of a difference of two
KL\ divergences, so we can write:%
\begin{equation}
\ell_{k+1}(\theta)-\ell_{k}(\theta)=N_{k+1}\left(  D_{KL}(\alpha_{\ast
}||\alpha_{k})-D_{KL}(\alpha_{\ast}||\theta)\right)  +.... \label{diffodiffo}%
\end{equation}
We now approximate the lefthand side. The small $N_{k}/N$ limit means we can
replace the finite difference on the lefthand side by a derivative. More
precisely, introduce a continuous parameter $\tau\in\lbrack0,1]$, which we can
partition up into discretized values $\tau_{k}$ with small timestep
$\delta\tau_{k}=\tau_{k+1}-\tau_{k}$ between each step:
\begin{equation}
\tau_{k}\equiv\frac{1}{N}(N_{1}+...+N_{k})\,\,\,\text{and}\,\,\,\delta\tau
_{k}\equiv\frac{N_{k}}{N}.
\end{equation}
So, instead of writing $\pi_{k}(\theta)$, we can instead speak of a
continuously evolving family of distributions $\pi(\tau;\theta)$. Similarly,
we write $\alpha\left(  \tau\right)  $ to indicate the continuous evolution
used in the parameter appearing in $f(e|\alpha_{k})=f(e|\alpha(\tau_{k}))$.
The finite difference can therefore be approximated as:
\begin{equation}
\ell_{k+1}(\theta)-\ell_{k}(\theta)=\ell^{\prime}(\theta;\tau_{k})\delta
\tau_{k}+\frac{1}{2}\ell^{\prime\prime}(\theta;\tau_{k})(\delta\tau_{k}%
)^{2}+...,
\end{equation}
where the prime indicates a partial derivative with respect to the timestep,
e.g. $\ell^{\prime}=\partial\ell/\partial\tau$. Working in the approximation
$N\gg N_{k}\gg1$, equation (\ref{diffodiffo}) is then given to leading order
by:
\begin{equation}
\frac{1}{N}\frac{\partial\ell(\theta;\tau)}{\partial\tau}=D_{KL}(\alpha_{\ast
}||\alpha(\tau))-D_{KL}(\alpha_{\ast}||\theta)+.... \label{integrodiffo}%
\end{equation}
To avoid overloading the notation, we write this as:%
\begin{equation}
\frac{1}{N}\frac{\partial\ell(\theta;\tau)}{\partial\tau}=D(\alpha
(\tau))-D(\theta)+....
\end{equation}
Observe that the $N_{k}$ dependence has actually dropped out from the
righthand side; it only depends on the total number of events $N$. In terms of
the posterior $\pi(\theta;\tau)=\exp\ell(\theta;\tau)$, we have:%
\begin{equation} \label{Posterior DiffEQ}
\frac{\partial\pi(\theta;\tau)}{\partial\tau}=\pi(\theta;\tau)\frac
{\partial\ell(\theta;\tau)}{\partial\tau}=N\pi(\theta;\tau)\left(
D(\alpha(\tau))-D(\theta)\right)  \text{.}%
\end{equation}
A formal solution to the posterior is then:%
\begin{equation} \label{Posterior Solution}
\pi(\theta;\tau)=\exp N\int^{\tau}d\tau^{\prime}\text{ }\left(
D(\alpha(\tau))-D(\theta)\right)  \text{.}%
\end{equation}

\subsection{Observable Flows}

Given an inference scheme over a random variable with parameters $\theta$, we regard an observable as a function of the parameters; i.e. in the classical sense with the parameter space serving as a phase space for the theory. Given such an observable, $O(\theta)$,
we now ask about the $\tau$ dependence, as obtained by evaluating the expectation value:%
\begin{equation}
\overline{O}(\tau)=\int d\theta\text{ }O(\theta)\pi(\theta;\tau).
\end{equation}
This is subject to a differential equation, as obtained by differentiating
both sides with respect to $\tau$:%
\begin{equation}
\frac{\partial\overline{O}(\tau)}{\partial\tau}=N\int d\theta\text{ }%
O(\theta)\pi(\theta;\tau)\left(  D(\alpha(\tau))-D(\theta)\right)  ,
\label{diffoO}%
\end{equation}
and using equation
(\ref{Posterior DiffEQ}). More compactly, we can write this as:%
\begin{equation}
\left(  \frac{\partial}{\partial\tau}-N(D(\alpha(\tau))-\overline{D}%
(\tau))\right)  \overline{O}(\tau)=-N\text{Var}(O,D), \label{connectionform}%
\end{equation}
where Var$(O,D)$ is just the variance between the operators $O$ and $D$:
\begin{equation} \label{OD Covariance}
\text{Var}(O,D)=\left\langle (O-\overline{O})(D-\overline{D})\right\rangle
_{\pi(\theta;\tau)}.
\end{equation}

Before proceeding, let us consider the trivial observable $O(\theta) = 1$. This observable determines the normalization condition imposed on $\pi(\theta;\tau)$ as a formal probability density. Using equation (\ref{diffoO}), we find:
\begin{equation}
    0 = N \int d\theta \, \pi(\theta;\tau) (D(\alpha(\tau)) - D(\theta)) = N(D(\alpha(\tau)) - \Bar{D}(\tau))
\end{equation}
which implies that:
\begin{equation} \label{Expectation of KL}
    D(\alpha(\tau)) = \overline{D}(\tau) \, .
\end{equation}
Thus, the equation obeyed by arbitrary observables is given by:
\begin{equation}
    \frac{\partial}{\partial \tau} \overline{O}(\tau) = - N\text{Var}(O,D) \, .
\end{equation}
To proceed further, it is helpful to work in terms of a rescaled time coordinate
$T \equiv \tau N$. In terms of this variable, our equation becomes:%
\begin{equation} \label{Observable Flow Equation}
\frac{\partial}{\partial T}
\overline{O}(T)= -\text{Var}(O,D) \, ,
\end{equation}
so that the $N$ dependence has dropped out. By expanding the covariance we can also write this equation as:
\begin{equation} \label{Most General Observable Flow Equation}
    \frac{\partial \overline{O}}{\partial T} = \overline{O} \; \overline{D} - \int d\theta \pi(\theta;T) O(\theta) D(\theta) \, .
\end{equation}
We now turn to the interpretation of this equation in various regimes.

\subsection{Generic Observables at Late T}

Our approach to analyzing this equation will begin with expanding the observables in a power series to obtain manageable expressions that have interpretations as governing late $T$ behavior. In particular, we will use the expansion of the divergence:
\begin{equation} \label{Taylor Expansion of KL}
    D(\theta) \approx \frac{1}{2} \mathcal{I}_{ij}\bigg\rvert_{\alpha_*} (\theta - \alpha_*)^i (\theta - \alpha_*)^j + \mathcal{O}(1/T^3) \, ,
\end{equation}
where $\mathcal{I}\rvert_{\alpha_*} = \mathcal{I}_*$ is the Fisher information metric evaluated at the true underlying mean of the parameter distribution. It is necessary to expand around this parameter value if one wishes to represent the KL-Divergence as a quadratic form with no constant or linear contribution -- that is, we have used the fact that $\alpha_*$ is the minimizing argument of $D(\theta)$ to set the constant and linear order terms in (\ref{Taylor Expansion of KL}) to zero. Then, at late times any arbitrary observable satisfies the equation:
\begin{flalign}
    \frac{\partial \overline{O}}{\partial T} = & \frac{1}{2} \mathcal{I}^*_{kl} \; \overline{O} \int d\theta \pi(\theta;T) (\theta - \alpha_*)^k (\theta - \alpha_*)^l  \nonumber \\
    & - \frac{1}{2} \mathcal{I}^*_{kl}\int d\theta \pi(\theta;T) O(\theta) (\theta - \alpha_*)^k (\theta - \alpha_*)^l + \mathcal{O}(1/T^3)   \, .
\end{flalign}

\subsection{Centralized Moments at Late T}

At this juncture, let us turn our attention to a special class of observables. Namely, those of the form:
\begin{equation}
    C^{i_1 ... i_{2l}}(\theta) = \prod_{j = 1}^{2l} (\theta - \alpha_*)^{i_j}  \, .
\end{equation}
Such observables are precisely the pre-integrated centralized moments of the $T$-posterior. More precisely, this is true at sufficiently late times in which $\alpha(T) \sim \alpha_*$, meaning the parameter distribution has centralized around its true mean. The observable flow equation for these observables therefore governs the $T$-dependence of the centralized moments:
\begin{equation}
    \overline{C}^{i_1 ... i_{2l}}(T) = \int d\theta \pi(\theta;T) \prod_{j=1}^{2l} (\theta - \alpha_*)^{i_j}
\end{equation}
which satisfy the differential equation:
\begin{multline}
    \frac{\partial \overline{C}^{i_1 ... i_{2l}}}{\partial T} = \frac{1}{2} \mathcal{I}^*_{kl} \overline{C}^{i_1 ... i_{2l}} \int d\theta \pi(\theta;T) (\theta - \alpha_*)^k (\theta - \alpha_*)^l \\
     - \frac{1}{2} \mathcal{I}^*_{kl} \int d\theta \pi(\theta;T) C^{i_1 ... i_{2l}}(\theta) (\theta - \alpha_*)^k (\theta - \alpha_*)^l + \mathcal{O}(1/T^3)   \, .
\end{multline}
Using our notation this can be written more briefly as:
\begin{equation}
    \frac{\partial}{\partial T} \overline{C}^{i_1 ... i_{2l}} = \frac{1}{2} \mathcal{I}^*_{kl} \overline{C}^{i_1 ... i_{2l}} \overline{C}^{kl} - \frac{1}{2} \mathcal{I}^*_{kl} \overline{C}^{i_1 ... i_{2l} \; kl} + \mathcal{O}(1/T^3)  \, .
\end{equation}

\subsection{Centralized Moments for Gaussian Distributions}

The analysis we have done up to this point is valid for the centralized moments of any arbitrary posterior distribution. Now, we will specialize to the case that the posterior distribution is Gaussian at late $T$. This is quite generic, and will always be the case when the parameters being inferred over are truly non-stochastic. The aspect of the Gaussian model which is especially useful is that we can implement Isserlis', a.k.a, Wick's theorem, to reduce $2l$-point functions into sums of products of $2$-point functions (i.e. the covariance). In particular, we have the formula:
\begin{equation} \label{Isserlis Theorem}
    \overline{C}^{i_1 ... i_{2l}} = \sum_{p \in \mathcal{P}_{2l}^2} \prod_{(r,s) \in p} \overline{C}^{i_r i_s}  \, .
\end{equation}
Here $\mathcal{P}_{2l}^2$ is the set of all partitions of $2l$ elements into pairs, and a generic element $p \in \mathcal{P}_{2l}^2$ has the form $p = \{(r_1, s_1) , ..., (r_l, s_l)\}$, hence the notation in the product. Isserlis' / Wick's theorem implies that for a Gaussian model it is sufficient to understand the $T$-dependent behavior of the $2$-point function, and the behavior of the other $2l$-point functions immediately follow suit. Therefore, let us consider the equation satisfied by $\overline{C}^{ij}$:
\begin{equation} \label{Covariance Equation}
    \frac{\partial}{\partial T} \overline{C}^{ij} = \frac{1}{2} \mathcal{I}^*_{kl} \overline{C}^{ij}\overline{C}^{kl} - \frac{1}{2} \mathcal{I}^*_{kl} \overline{C}^{ijkl} + \mathcal{O}(1/T^3) \, .
\end{equation}
Using (\ref{Isserlis Theorem}) we can write:
\begin{equation}
    \overline{C}^{ijkl} = \overline{C}^{ij} \overline{C}^{kl} + \overline{C}^{ik} \overline{C}^{jl} + \overline{C}^{il} \overline{C}^{jk}
\end{equation}
Plugging this back into equation (\ref{Covariance Equation}) we get:
\begin{flalign}
    \frac{\partial}{\partial T} \overline{C}^{ij} = & \frac{1}{2} \mathcal{I}^*_{kl} \overline{C}^{ij} \overline{C}^{kl} - \frac{1}{2} \mathcal{I}^*_{kl} \left(\overline{C}^{ij} \overline{C}^{kl} + \overline{C}^{ik} \overline{C}^{jl} + \overline{C}^{il} \overline{C}^{jk}\right) + \mathcal{O}(1/T^3) \\
    & = -\mathcal{I}^*_{kl} \overline{C}^{ik} \overline{C}^{jl} + \mathcal{O}(1/T^3)
\end{flalign}
where we have used the fact that $\mathcal{I}^*_{ij}$ is symmetric. The equation satisfied by the covariance of a Gaussian distribution at late $T$ is thus:
\begin{equation} \label{Cramer Rao Equation}
    \frac{\partial}{\partial T} \overline{C}^{ij} + \mathcal{I}^*_{kl} \overline{C}^{ik} \overline{C}^{jl} = 0 + \mathcal{O}(1/T^3) \, .
\end{equation}
At this stage, we can recognize our equation as predicting familiar behavior for the $2l$-point functions of a Gaussian posterior.

\subsection{Cram\'{e}r-Rao Solution}

Cram\'{e}r and Rao \cite{Rao} demonstrated  that there is a lower bound on the variance of any unbased estimator which is given by the inverse of the Fisher information. An estimator that saturates this bound is as efficient as possible and reaches the lowest possible mean squard error.

We now show that the evolution equation (\ref{Cramer Rao Equation}) is satisfied when the model saturates the Cram\'{e}r-Rao bound. That is, at sufficiently late $T$, we take the bound to be saturated such that:
\begin{equation} \label{Cramer-Rao}
    \overline{C}_{CR}^{ij} = \frac{\mathcal{I}_*^{ij}}{T} + \mathcal{O}(1/T^2) \, .
\end{equation}
One can see that (\ref{Cramer-Rao}) is then a solution to (\ref{Cramer Rao Equation}) by straightforward computation:
\begin{equation}
    \frac{\partial}{\partial T} \overline{C}_{CR}^{ij} = \frac{\partial}{\partial T} (\frac{\mathcal{I}_*^{ij}}{T}) = -\frac{\mathcal{I}_*^{ij}}{T^2} = -\mathcal{I}^*_{kl} \frac{\mathcal{I}_*^{ik}}{T} \frac{\mathcal{I}_*^{jl}}{T} = -\mathcal{I}^*_{kl} \overline{C}_{CR}^{ik} \overline{C}_{CR}^{jl} \, .
\end{equation}
We leave it implicit that there could be correction terms of higher order in the expansion $1/T$.
The $T$-dependent behavior of any arbitrary $2l$-point function in the theory is subsequently given by:
\begin{equation}
    \overline{C}_{CR}^{i_1 ... i_{2l}} = \frac{1}{T^l} \sum_{p \in \mathcal{P}_{2l}^2} \prod_{(r,s) \in p} \mathcal{I}_*^{i_r i_s}   \, .
\end{equation}

\subsection{Higher Order Effects}

The assumptions that led to the saturation of the Cram\'{e}r-Rao bound are based on two leading order approximations: Firstly, that the maximum likelihood parameter is near the ``true" value, and secondly, that the posterior distribution is approximately Gaussian. Both of these assumptions become increasingly valid at late update times ie. with more data, hence we have expressed our equations as a power series expansion in the small quantity, $1 / T$.

One can look for corrections to these assumptions by systematically reintroducing higher order effects via a perturbation series. To be precise, as one moves into earlier update times, there will be contributions to the KL-Divergence which are higher than quadratic order in $\theta$. Similarly, as one moves away from a Gaussian posterior, either by moving back in time or by including some additional implicit randomness to the parameters, one finds new terms in the posterior distribution away from Gaussianity which also lead to new terms in the Observable Flow equations.

We provide an outline of the perturbative analysis including effects both from the additional higher order expansion of the KL-Divergence and from the deviation of the Posterior from Gaussianity. To begin, the KL-Divergence can be Taylor expanded into a power series in the un-integrated $n$-point functions as follows:
\begin{equation}
    D(\theta) = \sum_{n = 2}^{\infty} \frac{1}{n!} \mathcal{I}^{(n)}_{i_1 ... i_n} C^{i_1 ... i_n}(\theta)
\end{equation}
where
\begin{equation}
    \mathcal{I}^{(n)}_{i_1 ... i_n} = \prod_{j = 1}^n \frac{\partial}{\partial \theta^{i_j}} D(\theta) \bigg\rvert_{\theta = \alpha_*} \, .
\end{equation}
We will now perturb the posterior distribution away from Gaussiantiy as follows:
\begin{equation}
    \pi(\theta;T) = \text{Gaussian} \cdot \; e^{-\lambda f(\theta)},
\end{equation}
where in the above, $f(\theta)$ is treated as an arbitrary bounded function, and the size of the small parameter $\lambda$ governing the perturbation is, in general, dependent on the update time $T$. We take the unperturbed Gaussian Distribution to be centered around the maximum likelihood estimate (MLE) with covariance given by the two-point correlator. The expectation value of an observable $O(\theta)$ can therefore be expanded in a series with respect to $\lambda$:
\begin{equation}
    \langle O \rangle_{\pi} = \langle e^{-\lambda f(\theta)} O(\theta) \rangle_{\mathrm{Gauss}} = \sum_{n = 0}^{\infty} \frac{(-1)^n \lambda^n}{n!} \langle f(\theta)^n O(\theta) \rangle_{\mathrm{Gauss}} \, .
\end{equation}
Now, one can input these series expansions back into (\ref{Most General Observable Flow Equation}) to obtain higher order corrections to the scaling behavior of any arbitrary observable expectation value. Doing so explicitly and terminating the perturbation series at order $N$ in powers of $\lambda$ and order $M$ in powers of $1 / T$ we find:

\begin{multline}
    \sum_{n = 0}^{N} \frac{(-1)^n \lambda^n}{n!} \frac{\partial}{\partial T} \langle f(\theta)^n O(\theta) \rangle_{\mathrm{Gauss}} \\
    = \sum_{n = 0}^N \frac{(-1)^n \lambda^n}{n!} \langle f(\theta)^n O(\theta) \rangle_{\mathrm{Gauss}}\sum_{n' = 0}^N \sum_{m = 2}^{M} \frac{(-1)^{n'} \lambda^{n'}}{n'!} \frac{1}{m!} \mathcal{I}^{(m)}_{i_1 ... i_m} \langle f(\theta)^{n'} C^{i_1 ... i_m}(\theta) \rangle_{\mathrm{Gauss}} \\
    - \sum_{n = 0}^N \sum_{m = 2}^{M} \frac{(-1)^n \lambda^n}{n!} \frac{1}{m!} \mathcal{I}^{(m)}_{i_1 ... i_m} \langle f(\theta)^n O(\theta)C^{i_1 ... i_m}(\theta) \rangle_{\mathrm{Gauss}} + \mathcal{O}(T^{-M}, \lambda^N) \, .
\end{multline}

We will work just to the next to leading order. To see the impact of these higher order effects, we performed two simple numerical experiments in which the Observable Flow for the two-point function can be solved exactly. In each case we consider only a single parameter being inferred upon during the Bayesian update.
\begin{enumerate}
    \item In the first numerical experiment, we consider a perturbation  in which we accept terms in the expansion of the KL-Divergence up to fourth order, but in which the posterior is assumed to remain approximately Gaussian. In this case, the Observable Flow equation for the second centralized moment, $\overline{C}^{(2)}$, becomes:
    \begin{equation}\label{eqn:Iago}
        \frac{\partial \overline{C}^{(2)}}{\partial T} = -\mathcal{I}^{(2)} (\overline{C}^{(2)})^2 - \frac{1}{2} \mathcal{I}^{(4)} (\overline{C}^{(2)})^3 \, .
    \end{equation}
We fix the $\mathcal{I}^{(2)}$ and $\mathcal{I}^{(4)}$ by hand and then use the above equation to solve for the time dependence of ${C}^{(2)}$. This is done using numerical methods, and subsequently fit to a power-law of the form
\begin{equation}\label{PL}
{C}^{(2)}=\frac{a}{T^b} + c \, .
\end{equation}

The resulting curve has a power in which $b < 1$, and typically in the range between $0.65$ and $1$ depending on the ratio between $\mathcal{I}^{(2)}$ and $\mathcal{I}^{(4)}$ (only the ratio matters). The results of this numerical experiment are given in table (\ref{Proximity Perturbation}).

\begin{table}[]
\centering
\begin{tabular}
[c]{|c|c|c|c|}\hline
$\mathcal{I}^{(4)}/\mathcal{I}^{(2)}$ & $a$ & $b$ & $c$\\\hline
$0.1$ & $0.96$ & $0.99$ & $-4.7\times10^{-6}$\\\hline
$0.2$ & $0.92$ & $0.98$ & $-2.5\times10^{-4}$\\\hline
$0.3$ & $0.88$ & $0.96$ & $-6.9\times10^{-4}$\\\hline
$0.4$ & $0.83$ & $0.94$ & $-1.3\times10^{-3}$\\\hline
$0.5$ & $0.78$ & $0.92$ & $-2.1\times10^{-3}$\\\hline
$0.6$ & $0.74$ & $0.89$ & $-2.9\times10^{-3}$\\\hline
$0.7$ & $0.70$ & $0.87$ & $-3.9\times10^{-3}$\\\hline
$0.8$ & $0.66$ & $0.84$ & $-4.9\times10^{-3}$\\\hline
$0.9$ & $0.62$ & $0.82$ & $-6.0\times10^{-3}$\\\hline
$1$ & $0.58$ & $0.79$ & $-7.1\times10^{-3}$\\\hline
$1.1$ & $0.55$ & $0.77$ & $-8.2\times10^{-3}$\\\hline
$1.2$ & $0.52$ & $0.74$ & $-9.3\times10^{-3}$\\\hline
$1.3$ & $0.49$ & $0.72$ & $-1.0\times10^{-2}$\\\hline
$1.4$ & $0.47$ & $0.69$ & $-1.2\times10^{-2}$\\\hline
$1.5$ & $0.44$ & $0.67$ & $-1.3\times10^{-2}$\\\hline
\end{tabular}
\caption{Result of the first numerical experiment involving perturbations to the KL divergence. This entails numerically solving equation (\ref{eqn:Iago}) by fitting to a power-law $a T^{-b} + c$, as in equation (\ref{PL}). In all cases, the $R^2$ value is $\sim 0.99 + O(10^{-3})$. As the ratio $\mathcal{I}^{(4)} / \mathcal{I}^{(2)}$ increases, we observe that the size of the constant offset increases in magnitude and the exponent $b$ in $T^{-b}$ decreases.}
\label{Proximity Perturbation}
\end{table}

    \item In the second numerical experiment, we consider perturbations away from Gaussianity in which we accept terms of order $\lambda$ with $f(\theta) = \theta^4$, but regard the KL-Divergence as sufficiently well approximated at quadratic order. In this case, the two-point function has the form:
    \begin{equation}
        \overline{C}^{(2)} = \overline{C}^{(2)}_{G} - 15 \lambda (\overline{C}^{(2)}_G)^3
    \end{equation}
    where $\overline{C}^{(2)}_G$ is the expectation of the second centralized moment with respect to the Gaussian distribution. The Gaussian two-point function subsequently satisfies the ODE:
    \begin{equation}\label{eqn:Cassio}
        \frac{\partial}{\partial T}\left(\overline{C}^{(2)}_G - 15\lambda (\overline{C}^{(2)}_G)^3\right) = \frac{1}{2}\mathcal{I}^{(2)}(\overline{C}^{(2)}_{G})^2 - 15\lambda \mathcal{I}^{(2)}(\overline{C}^{(2)}_G)^3 - \frac{1}{2} \mathcal{I}^{(2)}\left(3 (\overline{C}^{(2)}_G)^2 - 105\lambda (\overline{C}^{(2)}_G)^4\right) \, .
    \end{equation}
    Again, this equation can be solved using numerical methods, and fit to a power-law of the form (\ref{PL}). The resulting curves exhibit scaling in which $b < 1$. The size of $b$ is governed by the ratio of $\mathcal{I}^{(2)}$ and $\lambda$. The results of this experiment for various values of this ratio can be found in (\ref{Gaussianity Perturbation}).

\begin{table}[]
\centering
\begin{tabular}
[c]{|c|c|c|c|}\hline
$\lambda/\mathcal{I}^{(2)}$ & $a$ & $b$ & $c$  \\\hline
$0.01$ & $0.78$ & $0.89$ & $-3.9\times10^{-3}$ \\\hline
$0.02$ & $0.63$ & $0.80$ & $-7.7\times10^{-3}$ \\\hline
$0.03$ & $0.51$ & $0.71$ & $-1.1\times10^{-2}$ \\\hline
$0.04$ & $0.42$ & $0.63$ & $-1.7\times10^{-2}$ \\\hline
$0.05$ & $0.36$ & $0.55$ & $-2.1\times10^{-2}$ \\\hline
$0.06$ & $0.31$ & $0.49$ & $-2.6\times10^{-2}$ \\\hline
$0.07$ & $0.28$ & $0.44$ & $-3.0\times10^{-2}$ \\\hline
$0.08$ & $0.25$ & $0.40$ & $-3.5\times10^{-2}$ \\\hline
$0.09$ & $0.23$ & $0.36$ & $-3.9\times10^{-2}$ \\\hline
$0.1$ & $0.22$ & $0.32$ & $-4.3\times10^{-2}$  \\\hline
$0.11$ & $0.21$ & $0.29$ & $-4.8\times10^{-2}$ \\\hline
$0.12$ & $0.20$ & $0.26$ & $-5.3\times10^{-2}$ \\\hline
$0.13$ & $0.19$ & $0.24$ & $-5.9\times10^{-2}$ \\\hline
$0.14$ & $0.19$ & $0.21$ & $-6.5\times10^{-2}$ \\\hline
$0.15$ & $0.19$ & $0.19$ & $-7.2\times10^{-2}$ \\\hline
\end{tabular}
\caption{Result of the second numerical experiment involving perturbation away from Gaussianity. This involves numerical solutions to equation (\ref{eqn:Cassio}) obtained by fitting to a power-law $a T^{-b} + c$, as in equation (\ref{PL}). In all cases, the  $R^2$ value is $\sim 0.99 + O(10^{-3})$. As the ratio $\lambda / \mathcal{I}^{(2)}$ increases, we observe that the size of the constant offset increases in magnitude and the exponent $b$ in $T^{-b}$ decreases.}
\label{Gaussianity Perturbation}
\end{table}
\end{enumerate}

The upshot of these experiments, and of this section, is that higher order corrections to the Cram\'{e}r-Rao solution of the Observable Flow equation can be implemented systematically by considering a bi-perturbation series which takes into account both changes to the KL-Divergence due to the proximity of the MLE and the data generating parameter, and deviation of the posterior distribution from a Gaussian. The impact of these corrections are to decrease the steepness of the learning curve, in accord with expectations from the Cram\'{e}r-Rao Bound.

\subsection{Hidden Variables and the Breaking of the Cram\'{e}r-Rao Bound}

In the last section we showed that the Dynamical Bayesian Inference scheme respects the Cram\'{e}r-Rao Bound as an upper limit on the rate at which the two-point function can scale with respect to the update time. An analogous phenomenon occurs in conformal field theories, and is often referred to as the ``unitarity bound,'' (see \cite{Mack:1975je}) which controls the strength of correlations as a function of distance in the underlying spacetime.\footnote{For example, in a relativistic conformal field theory (CFT) in $D \geq 2$ spacetime dimensions, a scalar primary operator $O(x)$ of scaling dimension $\Delta$ will have two-point function: $\langle O^{\dag}(x) O(x) \rangle \sim 1 / \vert x \vert ^{2 \Delta}$, and $\Delta \geq \Delta_{0}$ specifies a unitarity bound which is saturated by a free scalar field, i.e. a Gaussian random field. In the context of the Cram\'{e}r-Rao bound, the limiting situation is again specified by the case of a Gaussian. The analogy is not perfect, however, because we do not have the same notion of spacetime locality in Bayesian inference, and the referencing to the Gaussian case is different ($\Delta > 1$ for a CFT, but powers 1 / $T^b$ for $b < 1$ in the case of Cram\'{e}r-Rao. It is, nevertheless, extremely suggestive, and the physical intuition about how to violate various unitarity bounds will indeed have a direct analogy in the statistical inference setting as well.}

\begin{figure}[t!]
\begin{center}
\includegraphics[scale = 0.5, trim = {0cm 5.0cm 0cm 5.0cm}]{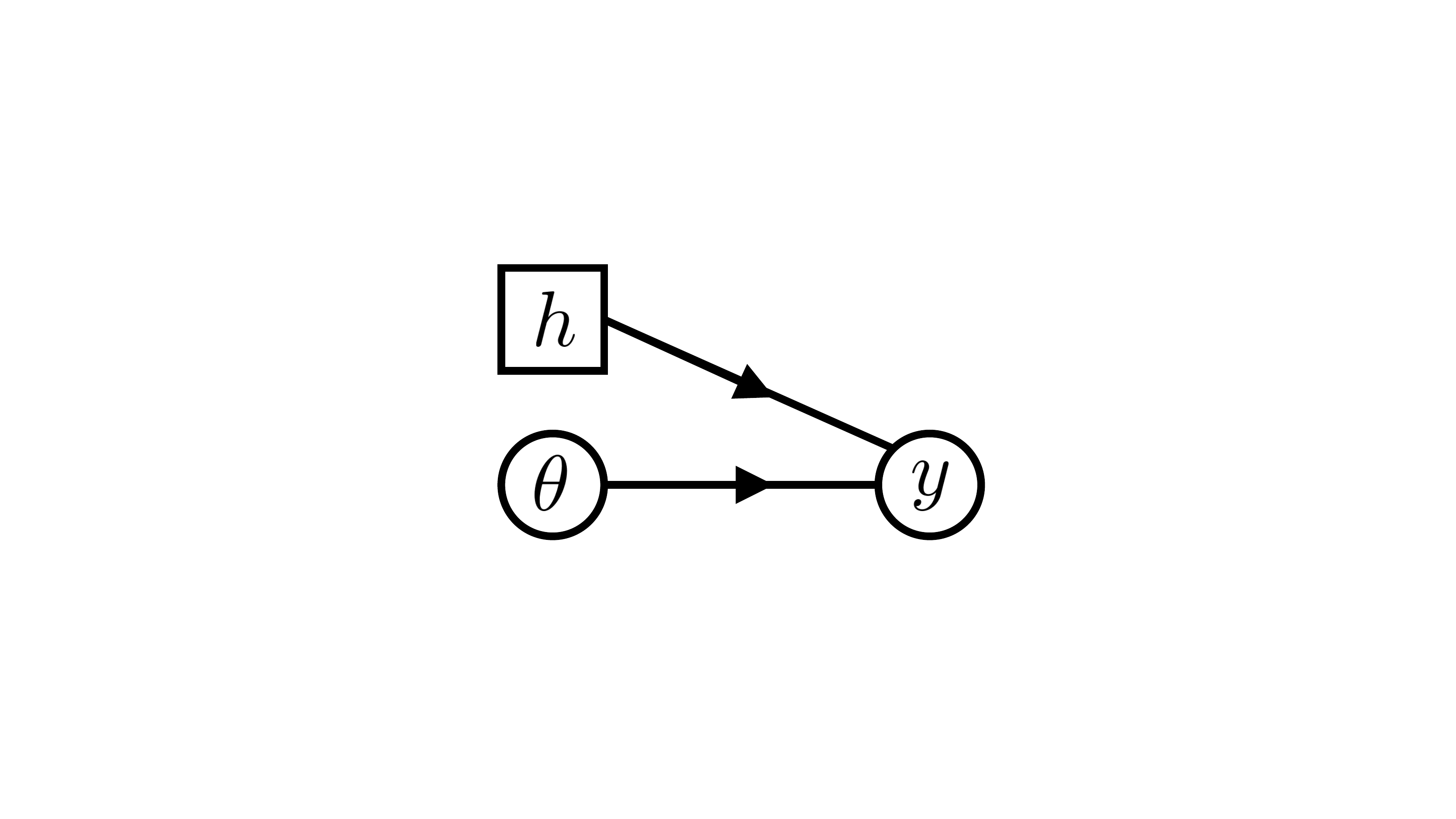}
\caption{Graphical model depiction of a conditional distribution $p(y|\theta,h)$
we wish to infer, where now we explicitly account for both visible training parameters $\theta$ and
hidden parameters $h$. The presence of such hidden variables can impact the inference scheme.}
\label{fig:HiddenVariables}
\end{center}
\end{figure}

Now, in the physical setting, a simple way to violate constraints from unitarity is to treat the system under consideration as ``open'', i.e. degrees of freedom can flow in or out. In the context of Bayesian inference, we have a direct analogy in terms of hidden variables $h$ which may also impact the distribution $f(y|\theta, h)$, but which we may not be able to access or even parameterize (see figure \ref{fig:HiddenVariables}). This can lead to dissipative phenomena, as well as driving phenomena.\footnote{Returning to the case of a CFT in $D$ dimensions, observe that a 4D free scalar can be modeled in terms of a collection of 3D scalars coupled along a discretized dimension. The unitarity bound for a scalar primary operator in 3D is $\Delta_{3D} \geq 1/2$, while in 4D it is $\Delta_{4D} \geq 1$.}

The basic setup is to consider a data generating distribution which belongs to a parametric family, $p(y \mid \theta,h)$ which depends on two sets of variables, $\theta$ (visible) and $h$ (hidden). An experimenter who is observing the data generated by this distribution may, either due to ignorance or by choice,\footnote{For example, a model builder may choose to fix the ``hidden" variables if they are close to their maximum likelihood values, or do not vary greatly across samples. In this regard the hidden variables may more aptly be identified as non-dynamical rather than hidden, but their impact is the same either way.} train a model that depends only on the variables, $\theta$, i.e. $f(y \mid \theta)$. Observables involving the hidden variables will evolve indirectly over the course of the update due to the changing of the total joint probability density over trained and hidden variables, however this evolution is not necessarily governed by a Bayesian updating scheme. Insofar as we have a reliable inference scheme at all, we can neglect the explicit time variation in the hidden variables, i.e., we can treat them as non-dynamical. Summarizing, only observables in the visible parameters $\theta$ will satisfy Observable Flow equations.

Now, although we are treating the hidden parameters as non-dynamical, they still enter in the KL-Divergence between the likelihood and the data generating model and therefore impact all observable flow equations. To be precise, at leading order
\begin{equation} \label{KL Hidden}
    D_{KL}(\Phi_* \parallel \Phi) = \frac{1}{2} \mathcal{I}_{AB} (\Phi - \Phi_*)^A (\phi - \Phi_*)^B  \, .
\end{equation}
Here we are using notation in which $\Phi = (\theta, h)$ is the complete set of parameters, and $\Phi_{\ast} = (\alpha_{\ast} , h_{\ast})$ (by abuse of notation) denotes the actual parameters from which we draw the distribution. The index $A = (i,I)$ spans all parameters, with the index $i = 1, ..., n$ corresponding to the trained parameters, and the index $I = 1, ... m$ corresponding to the hidden parameters. In this more explicit notation, the information metric appearing in (\ref{KL Hidden}) takes the form
\begin{equation}
    \mathcal{I} = \mathcal{I}_{ij} d\theta^i \otimes d\theta^j + \mathcal{I}_{IJ} dh^I \otimes dh^J + \mathcal{I}_{iI} d\theta^i \otimes dh^I + \mathcal{I}_{Ii} dh^I \otimes d\theta^i
\end{equation}
where $\mathcal{I}_{iI} = \mathcal{I}_{Ii}$.

Consider next the scaling of the two-point function between trained parameters. The scaling of the two-point function over the course of the Dynamical Bayesian inference scheme is dictated by the differential equation governing $C^{ij}(\theta)$. Using the observable flow equation, and remembering to include the complete KL-Divergence including contributions from both hidden and trained variables, one finds the equation:\footnote{Here we have extended the notation $\overline{C}^{A_1 ... A_n}$ to refer to the expectation value of $n$-point functions including arbitrary combinations of trained and hidden parameters:
\begin{equation}
    \overline{C}^{A_1 ... A_n} = \int d\theta \; dh \;  \rho(h \mid \theta) \pi(\theta;T) \prod_{i = 1}^n (\Phi - \Phi_*)^{A_i}
\end{equation}
$\rho(h \mid \theta)$ is a fixed conditional distribution encoding the probability density for hidden parameters given trained parameters.}
\begin{equation}
    \frac{\partial}{\partial T} \overline{C}^{ij} = -\mathcal{I}_{kl} \overline{C}^{ik} \overline{C}^{jl} - \mathcal{I}_{KL} \overline{C}^{iK} \overline{C}^{jL} - 2\mathcal{I}_{lL} \overline{C}^{il} \overline{C}^{jL} \, .
\end{equation}
The new contributions are the final two terms on the righthand side which depend on the covariance between trained and hidden variables. As noted above, such observables are to be considered as slowly varying in comparison with observables involving only trained parameters. Hence, for the purposes of this exercise we can regard these covariances as approximately constant in time.

We now make the well-motivated assumption that the joint probability model between the trained and hidden parameters is such that the covariance amongst all pairs of trained and hidden parameters satisfies the series of inequalities:
\begin{equation} \label{Inequalities}
    \overline{C}^{ij} \gg \overline{C}^{iI} \gg \overline{C}^{IJ}
\end{equation}
for all values of $i,j,I,J$. This assumption (along with the assumption that hidden variable observables are slowly varying) basically serve to justify the distinction between hidden and trained variables. If the hidden variables were rapidly varying and/or highly correlated with observed data it would not be reasonable to exclude them from the model. Alternatively, we may use these conditions as a criterion for \emph{defining} hidden variables as those variables which vary slowly and have limited covariance with relevant parameters.\footnote{This is again reminiscent of the splitting between fast and slow modes which one uses in the analysis of renormalization group flows.}
Under these assumptions, we can then write the ODE governing the scaling of the two-point function as:
\begin{equation} \label{Dynamical Bayes with Hidden Variables}
    \frac{\partial}{\partial T} \overline{C}^{ij} = -\mathcal{I}_{kl} \overline{C}^{ik} \overline{C}^{jl} - 2\mathcal{I}_{lL} \overline{C}^{il} \overline{C}^{jL}.
\end{equation}
From equation (\ref{Dynamical Bayes with Hidden Variables}) we can recognize that the evolution of the two-point function, (as well as the $n$-point functions) depends directly on the covariance between trained and hidden parameters. Note also that coupling to the hidden variables involves the off-diagonal terms of the information metric, $\mathcal{I}_{lL}$, which can in principle be either positive or negative, provided the whole metric is still positive definite. This can lead to a flow of information into the visible system (driving), or leakage out (dissipation).

The presence of this additional coupling to the hidden variables can produce an apparent violation of the Cram\'{e}r-Rao bound on just the visible sector. To see why, it is already enough to consider the simplest case where we have a single visible parameter $\theta$, with the rest viewed as hidden. In this case, the observable flow equation is:
\begin{equation} \label{Driven Observable Flow}
    \frac{\partial}{\partial T} \overline{C}^{11} = -\mathcal{I}_{11} (\overline{C}^{11})^2 - \beta \overline{C}^{11} + \mathcal{O}(\beta^2)
\end{equation}
where here
\begin{equation}
    \beta = 2 \mathcal{I}_{1I} \overline{C}^{1I}
\end{equation}
is twice the sum of the covariances of the trained parameter with the hidden parameters, and we have made it explicit that this is a leading order result in the size of these correlations. Dropping the order $\beta^2$ terms, the differential equation (\ref{Driven Observable Flow}) can be solved exactly:
\begin{equation}
\overline{C}^{11}(T) = \frac{\kappa \beta }{e^{\beta T} - \kappa \mathcal{I}_{11}},
\end{equation}
where the constant $\kappa$ depends on the initial conditions. Observe that for $\beta > 0$, $\overline{C}^{11}(T)$ decays exponentially at large $T$, i.e., faster than $1/T$. We interpret this as driving information into the visible sector. Conversely, for $\beta < 0$, we observe that the solution asymptotes to $-\beta / \mathcal{I}_{11} > 0$, i.e., we are well above the Cram\'{e}r-Rao bound (no falloff at large $T$ at all). We interpret this as dissipation: we are continually losing information.

In the above, we made several simplifying assumptions in order to analytically approximate the solution to the observable flow equations. At a phenomenological level, the interpolation from a simple $1/T$ behavior to an exponential decay law can be accomplished by a more general power-law of the form $1 / T^{1 + \nu}$ with $\nu > 0$, dependent on the particular inference scheme. This will be borne out by our numerical experiments, especially the ones in section \ref{sec:MNIST} involving inference in a neural network, where we study the loss function and its dependence on $T$.

It is interesting to note that the crossover between a power-law and exponential decay is also implicitly tied to the accuracy of the underlying model. This suggests that at lower accuracy there is more information left for the algorithm to draw into its estimates. As the accuracy improves, the available information decreases and hence the driven behavior is slowly deactivated, resulting in more approximately power-law type behavior. Stated in this way, it is interesting to ponder what the precise nature of this crossover is, and whether it may be regarded as a kind of phase transition. We leave a more fundamental explanation of this crossover behavior to future work.

\section{Dynamical Bayes for Gaussian Data}  \label{sec:GAUSS}

To give an analytic example of dynamical Bayesian updating, we now consider the illustrative case of sampling from a Gaussian distribution. An interesting special case is that of the Gaussian random process which can also be used to gain insight into the inference of neural networks (see, e.g., \cite{Neal}).

\subsection{Analysis for Multivariate Gaussian Data}

A $d$-dimensional Gaussian random variable can be regarded as a random variable distributed according to a family of distributions governed by two parameters -- a mean vector $\mu$, and a symmetric, positive semi-definite covariance matrix $\Sigma$. Explicitly:
\begin{equation} \label{Multivariate Gaussian}
    f(y \mid \mu, \Sigma) = ((2\pi)^d \det(\Sigma))^{-1/2} \; \exp(-\frac{1}{2} (y-\mu)^\intercal \Sigma^{-1} (y - \mu)) \, .
\end{equation}
Here $\Sigma^{-1}$ is the matrix inverse of the covariance; $\Sigma \Sigma^{-1} = \mathbb{I}$. By a simple counting argument, the number of free parameters governing the distribution of a $d$-dimensional Gaussian random variable is $d + \frac{d(d+1)}{2}$.

Bayesian inference over Gaussian data consists in determining a posterior distribution in the space of parameters $\Theta = (\mu, \Sigma)$. We can be slightly more general by allowing for reparameterizations of the space of parameters in terms of some $\theta \in \mathcal{S} \subset \mathbb{R}^{(d + \frac{d(d+1)}{2})}$, that is:
\begin{equation} \label{Reparameterization}
    \Theta = \Theta(\theta) = (\mu(\theta), \Sigma(\theta)) \, .
\end{equation}
Hence, the result of a Dynamical Bayesian inference procedure on Gaussian data is to determine a flow in the parameters, $\theta = \alpha(T)$, giving rise to a flow in the posterior distribution $\pi(\theta; T)$.

In the case of the Gaussian distribution, and many other standard distributions for that matter, we can say slightly more than what we could when the family governing data remains unspecified. In particular, we have an explicit form for the KL-Divergence between multivariate Gaussian distributions:
\begin{equation}
    D_{KL}((\mu_0, \Sigma_0) \parallel (\mu_1, \Sigma_1)) = \frac{1}{2}\left(\text{tr}(\Sigma_1^{-1} \Sigma_0) + (\mu_1 - \mu_0)^\intercal \Sigma_1^{-1} (\mu_1 - \mu_0) + \ln(\frac{\det(\Sigma_1)}{\det(\Sigma_0)}) - d \right) \, .
\end{equation}
This can be expressed in terms of $\theta_0$ and $\theta_1$ by composition with the reparameterization \ref{Reparameterization} provided $(\mu_a, \Sigma_a) = (\mu(\theta_a), \Sigma(\theta_a))$ for $a = 0, 1$:
\begin{equation}
D_{KL}(\theta_{0}\parallel\theta_{1})=\frac{1}{2}\left(
\begin{array}
[c]{l}%
\text{tr}(\Sigma(\theta_{1})^{-1}\Sigma(\theta_{0}))+\ln(\frac{\det
(\Sigma(\theta_{1}))}{\det(\Sigma(\theta_{0}))})-d)\\
+(\mu(\theta_{1})-\mu(\theta_{0}))^{\intercal}\Sigma(\theta_{1})^{-1}%
(\mu(\theta_{1})-\mu(\theta_{0}))
\end{array}
\right)  \,.
\end{equation}

Given the flowing of the parameters, $\alpha(T)$, and the true underlying parameters, $\alpha_*$, the posterior distribution is given by the solution to the Dynamical Bayesian updating equation:
\begin{equation}
    \pi(\theta; T) = \exp\left(\int_{0}^T dT' (D_{KL}(\alpha_* \parallel \alpha(T')) - D_{KL}(\alpha_* \parallel \theta)) \right) \, .
\end{equation}
This solution can be written in the form:
\begin{equation}
    \pi(\theta;T) = \exp\left(-T D_{KL} (\alpha_* \parallel \theta) \right) \exp\left(N \int_{0}^{T} dT' \, D_{KL}(\alpha_* \parallel \alpha(T')) \right)   \, .
\end{equation}
Note that the posterior distribution is proportional to the exponentiated KL-Divergence evaluated against the true underlying model parameter -- a standard result from the theory of large deviations:
\begin{equation}
    \pi(\theta;T) \propto \exp \left(-T D_{KL}(\alpha_* \parallel \theta)\right)   \, .
\end{equation}
Using the explicit form of the KL-Divergence for the normal distribution we find:
\begin{multline}
    \pi(\theta;T) \propto \det(\Sigma(\alpha_*)) \det(\Sigma(\theta))^{-1} \\
    \exp \left(T \left\{-\frac{1}{2}\text{tr}(\Sigma(\alpha_*)\Sigma(\theta)^{-1}) - \frac{1}{2} (\mu(\theta)-\mu(\alpha_*)) \Sigma(\theta)^{-1} (\mu(\theta)-\mu(\alpha_*))  \right\}\right)   \, .
\end{multline}
This distribution is of the form of a Normal-Inverse-Wishart with location parameter $\mu(\alpha_*)$ and inverse scale parameter $\Sigma(\alpha_*)$. This is precisely the expected result for the posterior of a normal data model whose conjugate prior distribution is Normal-Inverse-Wishart.

\subsection{Gaussian Random Processes}

Having addressed the Dynamical Bayesian inference of multivariate Gaussian data it becomes natural to discuss the Dynamical Bayesian inference of data which is distributed according to a Gaussian Random Process (GRP).\footnote{For an introduction to GRPs in machine learning, see reference \cite{GRP}.} A GRP may be interpreted as the functional analog of a Gaussian distribution. That is, instead of considering random vectors, one considers random \emph{functions}, and instead of specifying a mean vector and a covariance matrix one specifies a mean \emph{function} and a covariance \emph{kernel}. Let us be more precise:

Suppose the data we are interested in consists of the space of random functions, $\phi: D \rightarrow \mathbb{R}$.\footnote{Notice, this construction can be straightforwardly generalized to functions with values in arbitrary spaces, we consider maps into $\mathbb{R}$ for the sake of clarity.} To specify a GRP on such a sample space one must specify a mean function:
\begin{equation}
    \mu: D \rightarrow \mathbb{R}
\end{equation}
and a covariance kernel:
\begin{equation}
    \Sigma: D \times D \rightarrow \mathbb{R}
\end{equation}
Then, the distribution over functions takes the symbolic form:
\begin{equation} \label{Gaussian Random Process Distribution}
    f(\phi \mid \mu, \Sigma) = \mathcal{N} \exp \left( -\frac{1}{2} \int_{D \times D} dx dy \; (\phi(x)-\mu(x))\Sigma^{-1}(x,y)(\phi(y)-\mu(y)) \right) \, .
\end{equation}
Here $\Sigma^{-1}(x,y)$ is the inverse of $\Sigma(x,y)$ in the functional sense:
\begin{equation}
    \int_D dy \; \Sigma^{-1}(x,y) \Sigma(y,z) = \delta(x-z)
\end{equation}
and the prefactor $\mathcal{N}$ is formally infinite, and can be identified with the partition function (path integral) of the unnormalized GRP.

Taken literally, the distribution (\ref{Gaussian Random Process Distribution}) is difficult to use. It should rather be viewed as a set of instructions for how to interpret the GRP. Formally, a GRP is defined by restricting our attention to a finite partition of the domain $D$: $P = \{x_1, ..., x_n\} \subset D$. A functional random variable $f: D \rightarrow \mathbb{R}$ follows a Gaussian Process with mean $\mu(x)$ and covariance $\Sigma(x,y)$ if, for any such partition, the $n$-vector, $f_P = (f(x_1), ..., f(x_n))$ in a multivariate Gaussian random variable with mean $\mu = (\mu(x_1), ..., \mu(x_n))$ and covariance $\Sigma = \Sigma(x_i, x_j)$.

In this respect, the study of a GRP is precisely the same as the study of the multivariate Gaussian -- we need only restrict our attention to some finite partition of the domain of the functional random variable and then perform Dynamical Bayesian inference over the resulting multivariate normal random variable.

\section{Inference in the Ising Model} \label{sec:ISING}

We now turn to some numerical experiments to test the general framework of dynamical Bayesian updating. Along these lines, we consider the basic physical question: Given a collection of experimental data, how well can an observer reconstruct the underlying model?\footnote{See also \cite{Heckman:2013kza, Balasubramanian:2014bfa, Balasubramanian:2020lux, Fowler:2021oje} for related discussions.} To make this tractable, we assume that the particular physical model is known, but the couplings are unknown. A tractable example of this sort is the statistical mechanics of the Ising model, as specified by a collection of spins $\sigma = \pm 1$ arranged on a graph. In this setting, the statistical mechanics provides us with a probability distribution over spin configurations $\{ \sigma \}$ as specified by the Boltzmann factor:
\begin{equation}
P[\{\sigma \} | J] = \frac{1}{\mathcal{Z}(J)} \exp ( - H_{\mathrm{Ising}}[\{ \sigma \} | J]),
\end{equation}
where $\mathcal{Z}(J)$ is a normalization constant (i.e., the partition function) introduced to ensure a normalized distribution and $H_{\mathrm{Ising}}$ is the Ising model Hamiltonian with coupling constant $J$:
\begin{equation}
H_{\mathrm{Ising}}[\{ \sigma \} | J] = - J \underset{n.n.}{\sum} \sigma \sigma^{\prime}.
\end{equation}
In the above, the sum is over nearest neighbors on the graph. One can generalize this model in various ways, by changing the strength of any given bond in the graph, but for ease of analysis we focus on the simplest non-trivial case as stated here. In this case, each draw from the distribution $P[\{\sigma \} | J]$ is specified by a collection of spins $\{ \sigma \}$. We can bin all of these events, as we already explained in section \ref{sec:BAYES}, and this specifies a posterior distribution $\pi_{\mathrm{post}}(J ; T)$. Using this, we can extract the $T$ dependence of various observables, for example:
\begin{equation}\label{eqn:Jmoments}
\langle J^{m} \rangle = \int d J \, \pi_{\mathrm{post}}(J; T) J^{m}.
\end{equation}
We can also introduce the centralized moments:
\begin{equation}
\overline{C}^{m} = \langle (J - \langle J \rangle)^m \rangle.
\end{equation}

\begin{figure}[t!]
\begin{center}
\includegraphics[scale = 0.2, trim = {0cm 0.0cm 0cm 0.0cm}]{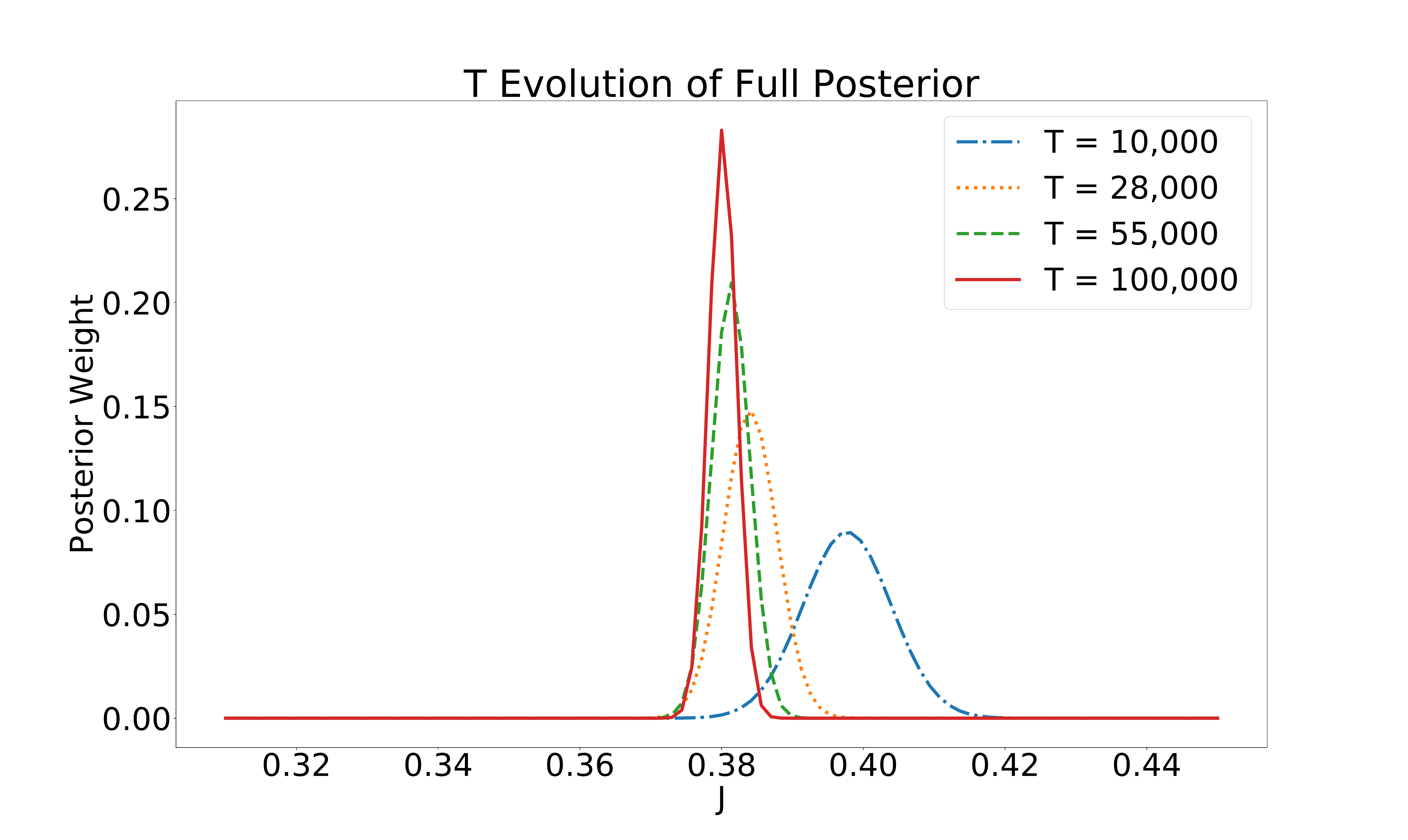}
\caption{Example of a trial in which the the posterior distribution over couplings is inferred at different update ``times'' incremented in steps of $900$ starting from an initial training at $T = 10,000$. We observe that the central value of the distribution converges to $J_{\ast} = 0.38$, and the width of the distribution narrows sharply. The match on higher order moments is displayed in table \ref{Ising Experiment Summary}.}
\label{fig:TEvoPost}
\end{center}
\end{figure}

\subsection{Numerical Experiment: 1D Ising Model}

As an explicit example, we now turn to the specific case of the 1D Ising model, i.e., a one-dimensional periodic lattice of evenly spaced spin. The Hamiltonian in this case is:
\begin{equation}
H_{\mathrm{Ising}} = - J \underset{1 \leq i \leq L}{\sum} \sigma_{i} \sigma_{i+1},
\end{equation}
with $\sigma_{L+1} \equiv \sigma_1$. We have an analytic expression for the partition function (see, e.g., \cite{Pathria:1996hda}),
and can also explicitly extract the Fisher information metric:
\begin{equation}
\mathcal{I}(J) = (L - 1)  \sech^{2}(J).
\end{equation}

\begin{figure}[t!]
  \begin{subfigure}[t]{.5\textwidth}
    \centering
    \includegraphics[width=\linewidth]{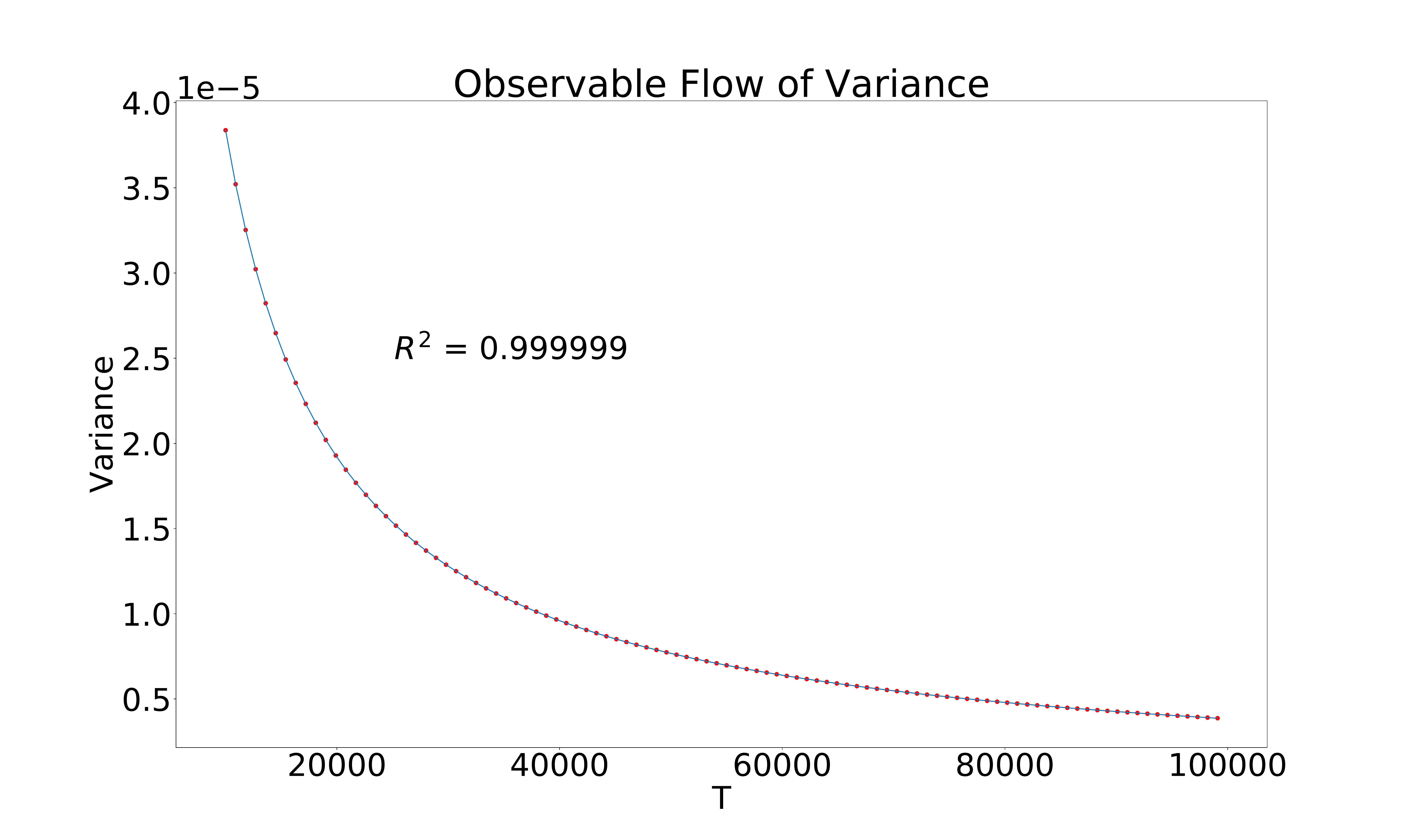}
    \caption{\textbf{Variance of Posterior Distribution}}
  \end{subfigure}
  \hfill
  \begin{subfigure}[t]{.5\textwidth}
    \centering
    \includegraphics[width=\linewidth]{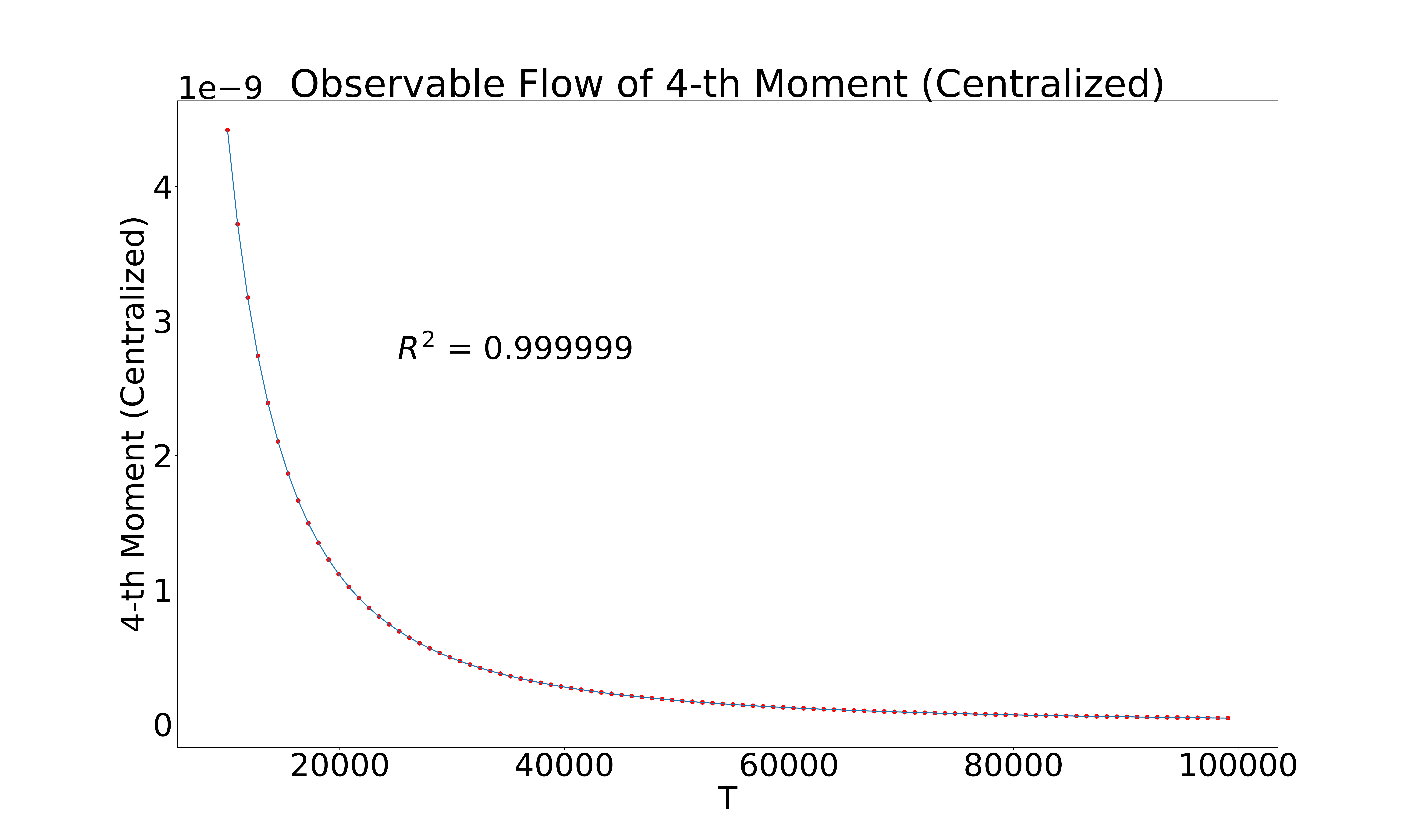}
    \caption{\textbf{Fourth Centralized Moment}}
  \end{subfigure}

  \medskip

  \begin{subfigure}[t]{.5\textwidth}
    \centering
    \includegraphics[width=\linewidth]{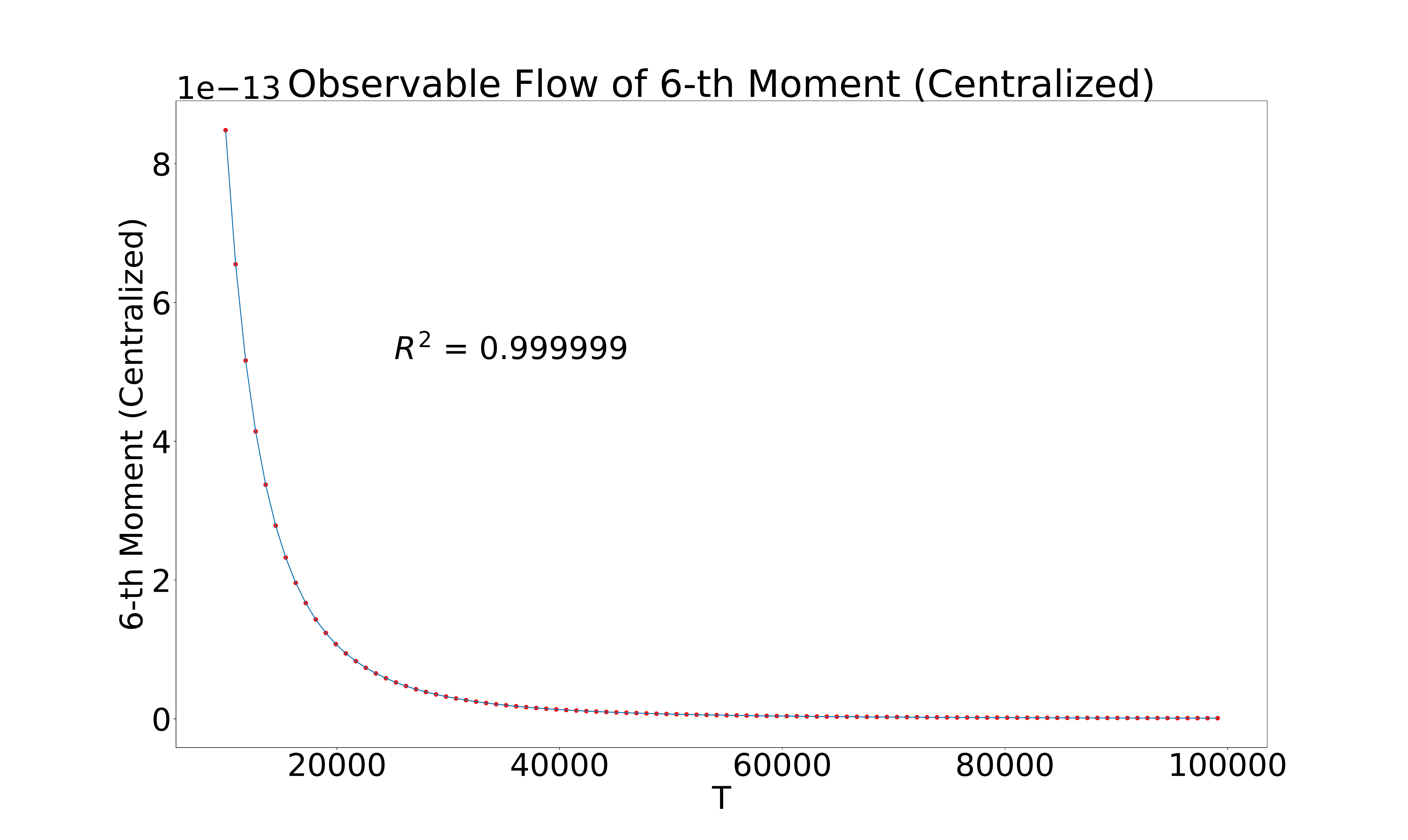}
    \caption{\textbf{Sixth Centralized Moment}}
  \end{subfigure}
  \hfill
  \begin{subfigure}[t]{.5\textwidth}
    \centering
    \includegraphics[width=\linewidth]{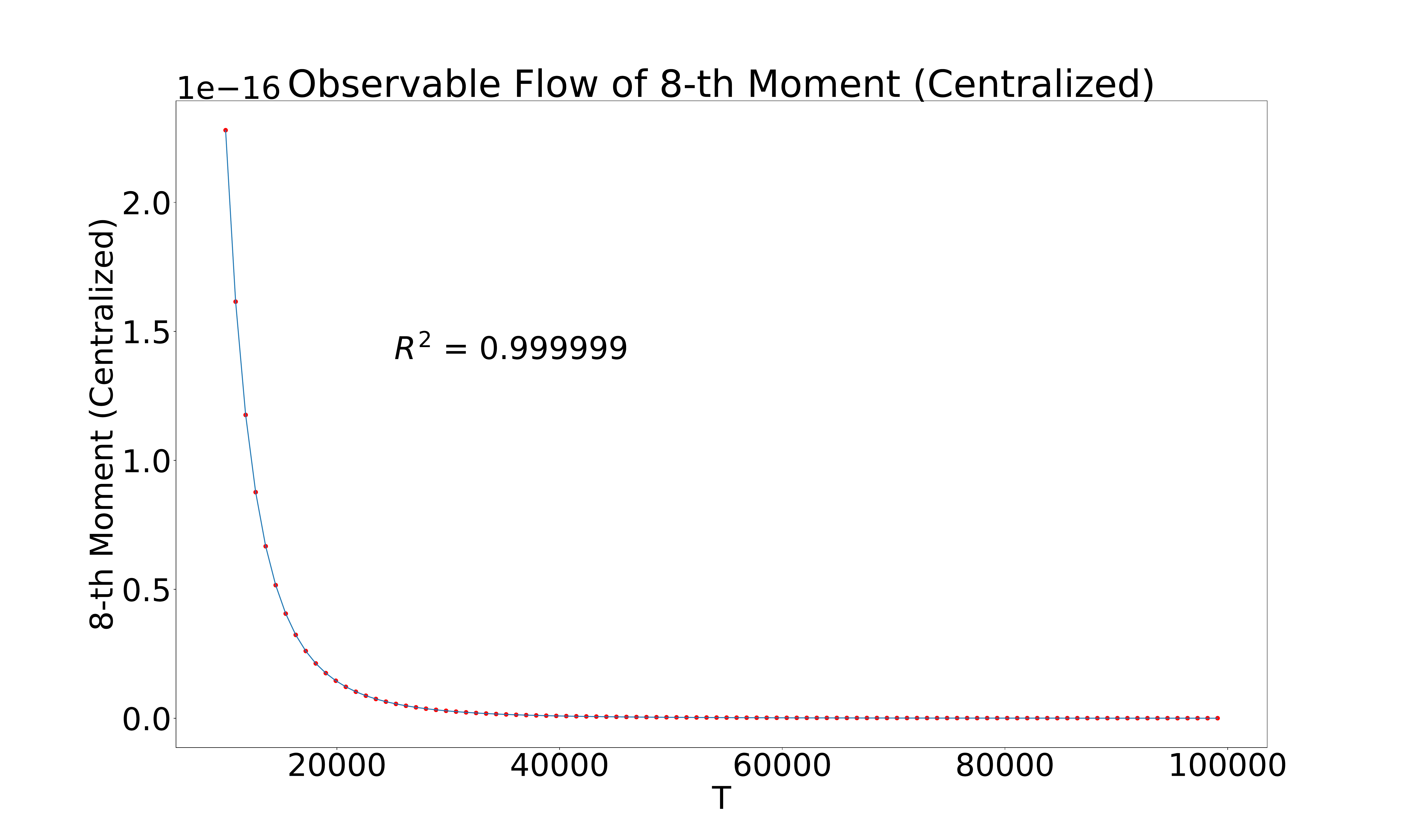}
    \caption{\textbf{Eighth Centralized Moment}}
  \end{subfigure}
  \caption{Observable flows for the first four centralized moments $\langle (J - \langle J \rangle)^{2l} \rangle$ for $l = 1,2,3,4$
  of the posterior distribution for the Ising Model Experiment. In all cases, we observe a power-law decay which is in close accord with the behavior saturated by the Cram\'{e}r-Rao bound (see equation \ref{Jcorr}).}
  \label{Ising Experiment Figures}
\end{figure}

\begin{figure}[h!]
    \centering
    \includegraphics[scale = .15]{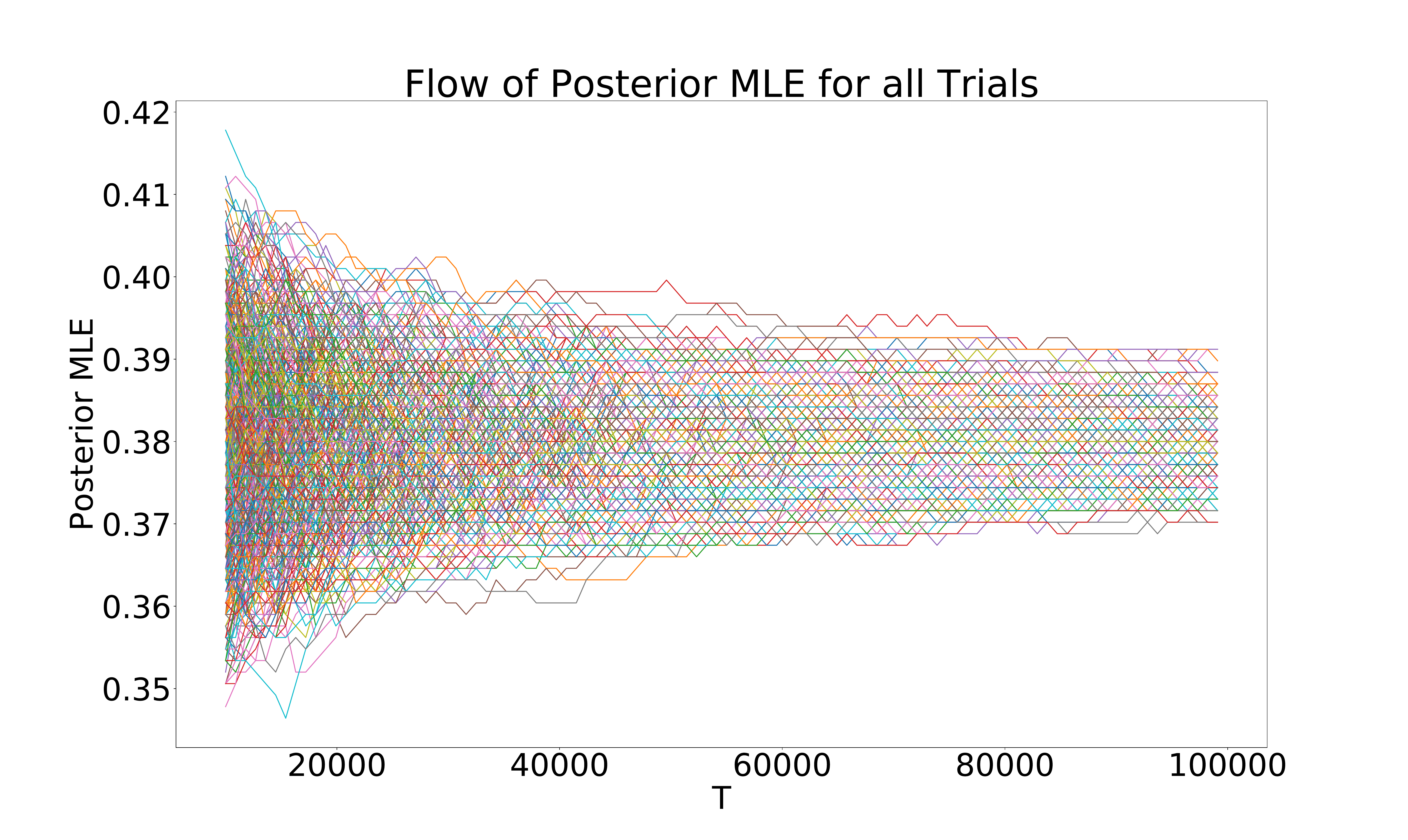}
    \caption{Dynamical Bayesian Trajectories for 1000 Ising Trials.}
    \label{1000 Trajectories}
\end{figure}

We would like to understand the convergence of the model to the true value of the parameter. Since the main element of our analysis involves adjusting the posterior distribution, it is enough to work with a small number of spins, i.e., $L = 4$. We take a benchmark value of $J_{\ast} = 0.38$ (so the Fisher information metric is  $\mathcal{I}(J_{\ast}) = 2.60533$) and track the dynamical Bayesian updating on the inference of this coupling. For a given trial, we performed a Bayesian update to track how well we could infer the value of the coupling constant. In each trial, we sampled from the Boltzmann distribution $100,000$ distinct spin configurations. Starting from the initial prior $J = 0$ (uniform distribution), we performed an initial update using $10,000$ events to get the first estimate for $J$. We then used the remaining $90,000$ events to obtain a series of sequential updates. The posterior was updated after the inclusion of every additional set of $900$ events. This then ran for a total of $1000$ time steps.

\begin{table}[t!]
\centering
\begin{tabular}
[c]{|c|c|c|}\hline
Moment & C-R Limit & Experiment\\\hline
$\left\langle \left(  J-\langle J\rangle\right)  ^{2}\right\rangle $ &
$0.38/T$ & $0.38/T^{0.9997}$\\\hline
$\left\langle \left(  J-\langle J\rangle\right)  ^{4}\right\rangle $ &
$0.44/T^{2}$ & $0.44/T^{1.9996}$\\\hline
$\left\langle \left(  J-\langle J\rangle\right)  ^{6}\right\rangle $ &
$0.84/T^{3}$ & $0.84/T^{2.9993}$\\\hline
$\left\langle \left(  J-\langle J\rangle\right)  ^{8}\right\rangle $ &
$2.28/T^{4}$ & $2.26/T^{3.999}$\\\hline
\end{tabular}
\caption{Comparison of predicted scaling for $n$-point functions from Dynamical Bayesian Inference in the limit where the Cram\'{e}r-Rao bound is saturated (see equation \ref{Jcorr}), and the observed scaling from the Ising Model Experiment. We have displayed additional significant figures to exhibit the extent of this match. Observe that in all cases, the experimentally determined power-law is of the form $1 / T^{1 - \nu}$ for $\nu > 0$, i.e., it respects the lower limit expected from the Cram\'{e}r-Rao bound.}
\label{Ising Experiment Summary}
\end{table}

For each trial we observe some amount of random fluctuation, but after averaging over $1000$ trials, we observe strikingly regular behavior, especially in the moments of the coupling $J$ as computed by the posterior distribution (see equation (\ref{eqn:Jmoments})). The late $T$ posterior distribution is Gaussian, and can be seen for a sample run at progressively later times in figure (\ref{fig:TEvoPost}). The observable flow of the even centralized moments for the update dependent posterior distribution can be seen below. Assuming we saturate the Cram\'{e}r-Rao bound, we find:
\begin{equation}\label{Jcorr}
    \langle (J - \langle J \rangle)^{2l} \rangle = \overline{C}^{2l} = \frac{(2l-1)!!}{(\mathcal{I}_*)^{l}} T^{-l}
\end{equation}
Where $n!! = \prod_{k = 0}^{[\frac{n}{2}]-1} (n-2k)$. This agrees very well with the numerical experiment,
as can be seen in figure (\ref{Ising Experiment Figures}) and summarized in table (\ref{Ising Experiment Summary}).

Finally, we note there is some statistical variation present on the space of trajectories for the maximum likelihood estimate (MLE) (see figure \ref{1000 Trajectories}). This makes manifest that there is statistical variation in any individual inference scheme, but that on aggregate, the paths converge to the maximum likelihood estimate. This observation inspires a path integral interpretation of dynamical Bayesian updating that we leave for future work.

\newpage

\section{Neural Networks and Learning} \label{sec:MNIST}

The Bayesian approach to neural networks was pioneered by Neal in \cite{Neal}. In what follows we will examine whether the dynamical inference model described in the present work can be applied to neural networks. We will take the viewpoint that a neural network is simply a model whose parameters are given by its weights and biases. Training a neural network using data infers the most likely set of weights given the training set (at least one hopes that this is true). As such one may adopt the view that the training of neural networks is a Bayesian problem of inferring a posterior distribution over the weights given the data available and then one chooses a net with the most likely weights from the posterior distribution. Note that training a network is a stochastic process where the outcome depends on the initialization of weights and the path taken through training.

To apply the reasoning in the paper we will examine how the trained neural network is dependent on the quantity of data used in its training. In particular we will measure how a trained neural network changes as we increment the amount of data used in the training process. We will certainly not be able to follow in a fully quantitative way the calculations in the previous sections because a neural network has far too many parameters (its weights) to carry out the Bayesian analysis explicitly. Instead we will empirically investigate whether the neural network follows a similar qualitative dependence on data as indicated by dynamical Bayesian updating. Insofar as the loss function can be approximated near the final inference in terms of quantities which are quadratic in the underlying $\theta$ parameters, we expect a simple power-law behavior as we approach a high level of accuracy. We expect the loss function to exhibit an exponential decaying profile when the inference is only moderately successful. The fact that we empirically observe precisely this sort of behavior provides support for the general picture developed in section \ref{sec:BAYES}.

Let us outline the experiment. For a helpful glossary of terms and additional background, see e.g. reference \cite{Roberts:2021fes}. The basic idea is that we will consider training a neural network using differing sample sizes from the same data set and see how loss depends on the amount of data. (For comparison we will repeat the whole experiment using the MNIST, Fashion-MNIST and CIFAR10 data sets.) The first neural network we use will have a very simple feedforward (FF) architecture. The input layer is a $28 \times 28$ layer, corresponding to the MNIST input data. Next is a simple 128 node dense  layer followed by the final 10 node output layer with softmax activation. The cost function is taken to be the categorical cross-entropy.

We also consider some experiments involving more sophisticated convolutional neural networks, training on the MNIST data set and the CIFAR10 data set. In the case of the MNIST data set, we consider a convolutional layer with kernel size 2 and filter size 64, followed by max pooling (with pool size 2), followed by a drop out layer (with drop out parameter 0.3) and then another convolution layer, kernel size 2 and filter size 32, then max pooling (with pool size 2), a dropout layer (parameter 0.3), followed by a dense layer with 256 neurons with rectified linear unit (ReLU) activation and a final dropout layer (parameter 0.5) and a final dense layer with 10 outputs and softmax activation.

For the CIFAR10 data we used a convolutional neural network with 3 convolutional layers with respective filter sizes  32, 64, and 128, with kernel size 3 for each layer, a max pooling layer with $3 \times 3$ poolsize was included after each convolutional layer. This set of convolution/pooling layers are then followed by a 128 node dense layer with ReLU activation followed by a dropout layer with dropout parameter 0.4 leading on to the final dense layer of 10 outputs with softmax activation.

The main difference between the convolutional neural networks used in the MNIST and CIFAR10 experiments, apart from having the larger input layer for CIFAR10 is the kernel size of the convolutions. In all cases the hyperparameters such as for dropout were untuned. Given that such hyperparameter tuning tends to depend on the specifics of the data being learnt for the purposes of the questions in this paper we did not consider hyperparameter tuning as necessary.

Crucially, we wish to investigate the dependence of the loss on the amount of data and not the amount of training of the network. Usually in training a neural network these two become connected since in any given epoch the amount of training depends on the amont of data. But crucially, neural networks often learn by repeated training using the same data set over many epochs. We are interested in the final state of the neural network after we have completed training.

We wish to keep the amount of training fixed and only compare the loss with different amounts of data used to do the training. (By training, we really mean the attempt to minimize the cost through some form of repeated gradient flow.)
To do this we link the number of epochs to the size of the training set we use. We have chosen to train over 4 epochs if the data set is maximal, i.e., 60,000 samples. This is a reasonable choice that produces good accuracy without overfitting. To demonstrate the reasoning behind this, consider training one neural net with $N$ data samples and another with $2N$. One training epoch for the network trained with $2N$ samples will have effectively twice the amount of training as the network with just $N$ samples. Thus to compare the effect of the larger data set as opposed to the amount of training we should train the network that uses the $2N$ data half the number of epochs as the one using the $N$ data set.

We train the networks using the \texttt{Adam} optimizer \cite{2014arXiv1412.6980K} with learning rate set to a standard 0.001. (For the full 60,000 samples and 4 epochs, this gives a healthy sparse categorical accuracy of around 0.97 for MNIST with the simple neural net.) After the network has been trained using the training set of $N$ samples it is tested on the full test set of 10,000 samples.

In what follows, we begin with a large sample size (e.g., 3,000) and then examine the loss after the training is complete as a function of the size of the training data set. We will then increase the training set size $N$ by some increment $\delta N$ where typically we take $\delta N$ to be around 500 and then repeat this until we reach a final data set an order of magnitude bigger e.g., 30,000 data points.
We then fit the resulting curve to the power-law behavior as expected from the dynamical Bayesian updating analysis. We find that for MNIST with the convolutional net the power-law is close to one but for Fashion-MNIST where the loss is higher, the power-law is of the form $1 / T^{1 + \nu}$ for $\nu > 0$. This is compatible with the contribution from hidden variables for Bayesian flows given in section \ref{sec:BAYES}.

We then repeat this with the CIFAR10 data set and the even more involved convolutional network where we find the exponential decay is a better fit than power-law indicating that the network has untrained parameters as in the hidden variable example discussed before.

The reader familiar with Stochastic and Batch gradient descent may feel that we are just doing the same thing in this experiment and these are just the traditional learning curves. This is \textit{not} the case since we train for multiple epochs and the curves measure only the loss as a function of \textit{total} data used in the training.

All the code is available to view in a Google Colab:

\url{https://colab.research.google.com/drive/1zNxHj7qCoE1-WzawqRTbaa9QhFWpQZqr?usp=sharing}

\subsection{Results}

Training neural nets is notoriously stochastic. To take this into account we actually perform multiple trials of each experiment (with different initial conditions in each case). We plot loss against $T$ and then fit to a power-law in each case. Performing multiple trials, we also extract the mean and variance for these fitting parameters, in particular the exponent appearing in the power-law fit. We also quote the root mean variance as an indicator of how robust the results are. For $10$ trials, the root variance of the power-law was between $8\%$ and $10\%$ depending on the data set in question.

\begin{figure}[h!]
  \begin{subfigure}[t]{.5\textwidth}
    \centering
    \includegraphics[width=\linewidth]{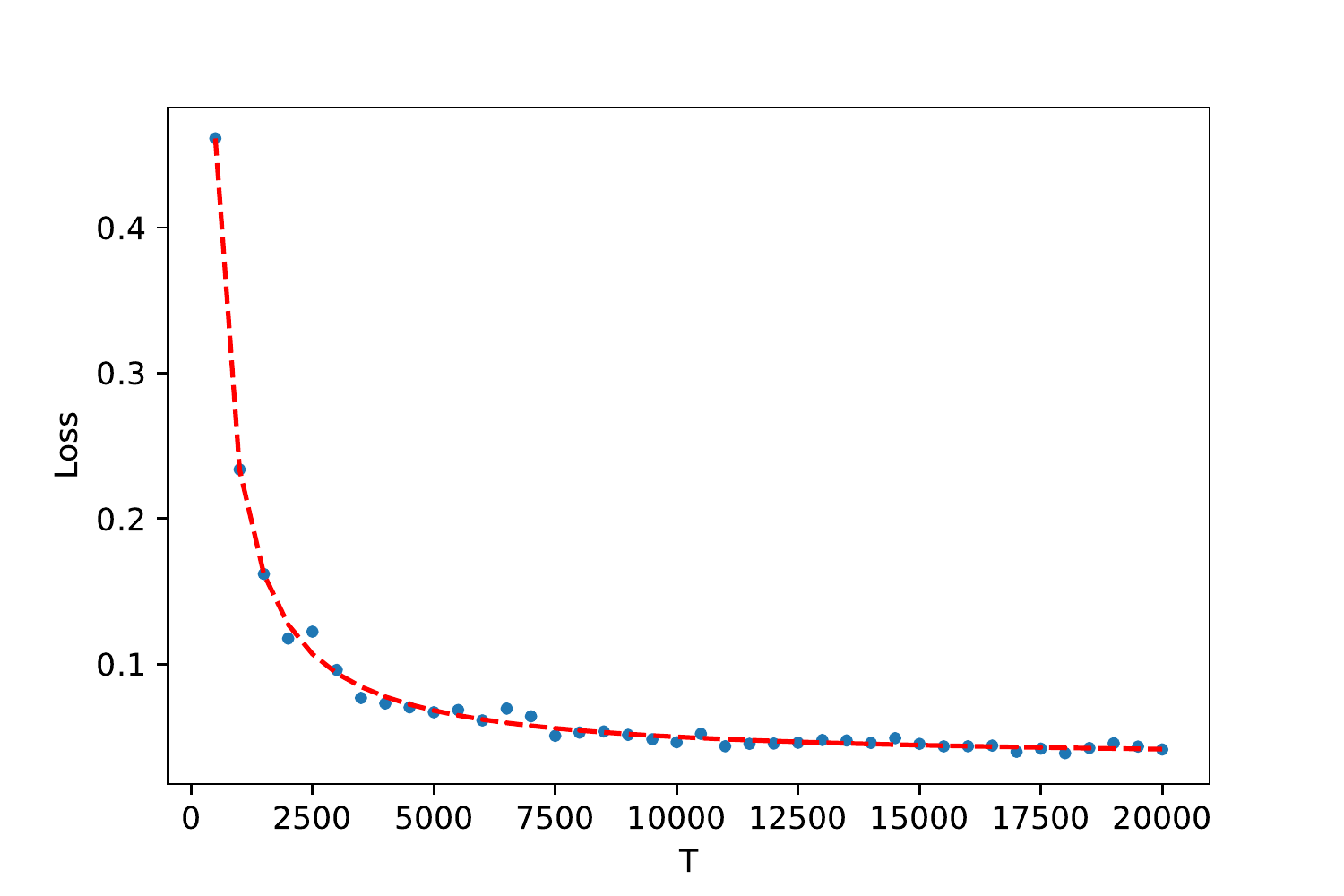}
    \caption{\textbf{MNIST Trial}}
  \end{subfigure}
  \hfill
  \begin{subfigure}[t]{.5\textwidth}
    \centering
    \includegraphics[width=\linewidth]{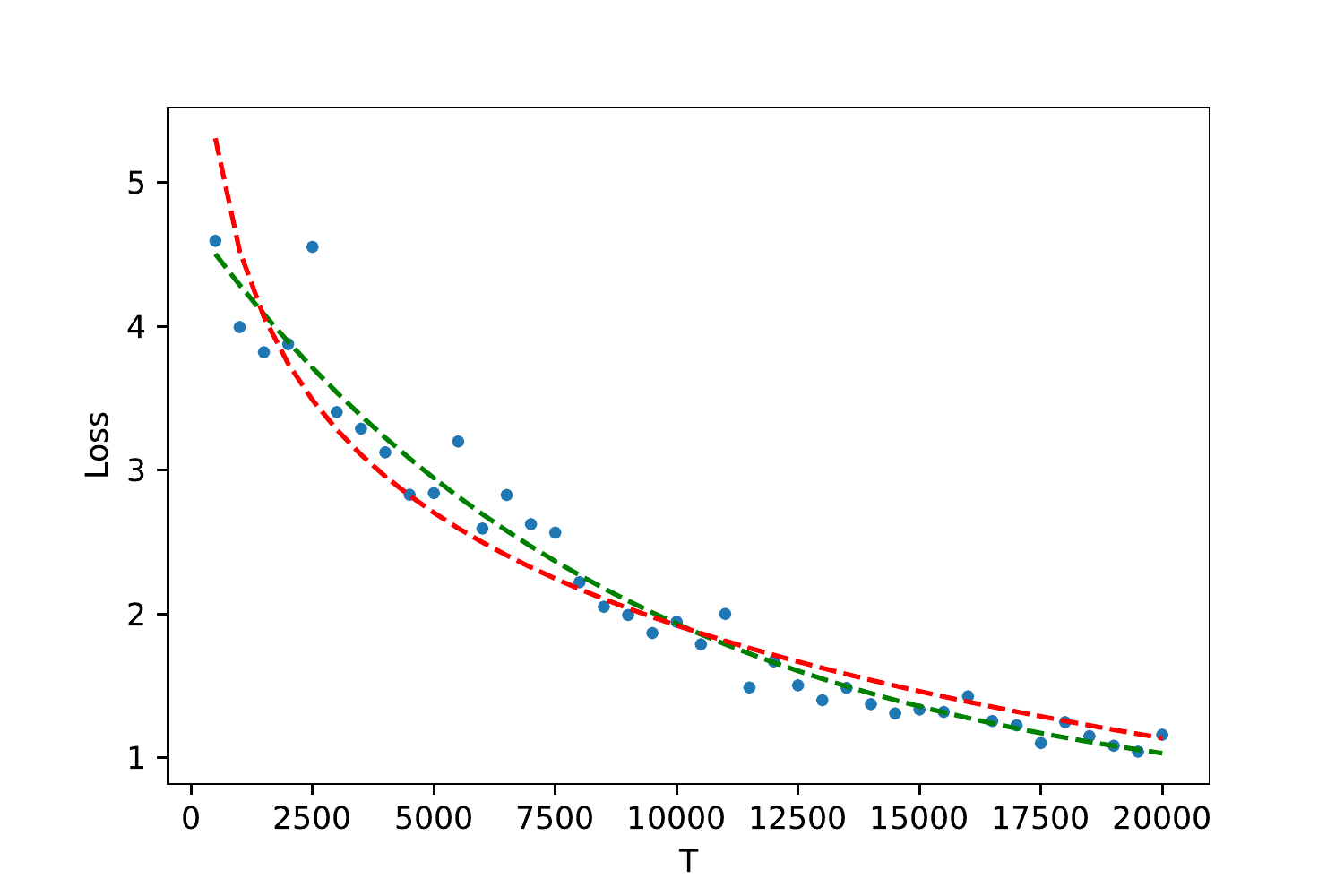}
    \caption{\textbf{CIFAR10 Trial}}
  \end{subfigure}
  \caption{Categorical cross-entropy loss as a function of $T$ in a convolutional neural network with varying amounts of trial data. Here, we display the results for a single complete run in the case of the MNIST and CIFAR10 data sets. In this case, we obtain a good fit to a power-law decay in the case of the MNIST data set, while in the case of the CIFAR10 data set, the lower accuracy is better fit by an exponential function (red curve) as opposed to a power-law (green curve). See also table \ref{tab:NET}. We collect the central values and variance of the decay law parameters for the different data sets in table \ref{tab:AGGRO}.}
  \label{CNNExperiments}
\end{figure}

We display here some representative examples of this analysis, as in figure \ref{FeedForwardExperiments} for the experiments with a feedforward neural network trained on the MNIST and Fashion-MNIST data sets, as well as figure \ref{CNNExperiments} for the convolutional neural network experiments trained on the MNIST and CIFAR10 data sets. In these plots we display the loss function (i.e., the categorical cross-entropy) on the vertical axis and the number of data samples used for training on the horizontal axis. In each case, we also display the corresponding fit for these particular examples, and the results are collected in table \ref{tab:NET}. As discussed above, an important aspect of these individual fits is that the actual parameters deviate from trial to trial; and so we also give the central values of the fitting parameters and their $1\sigma$ deviations. The mean values of the fitting parameters are displayed in table \ref{tab:AGGRO}.

\subsubsection{A Simple Feedforward Network}

\begin{figure}[t!]
  \begin{subfigure}[t]{.5\textwidth}
    \centering
    \includegraphics[width=\linewidth]{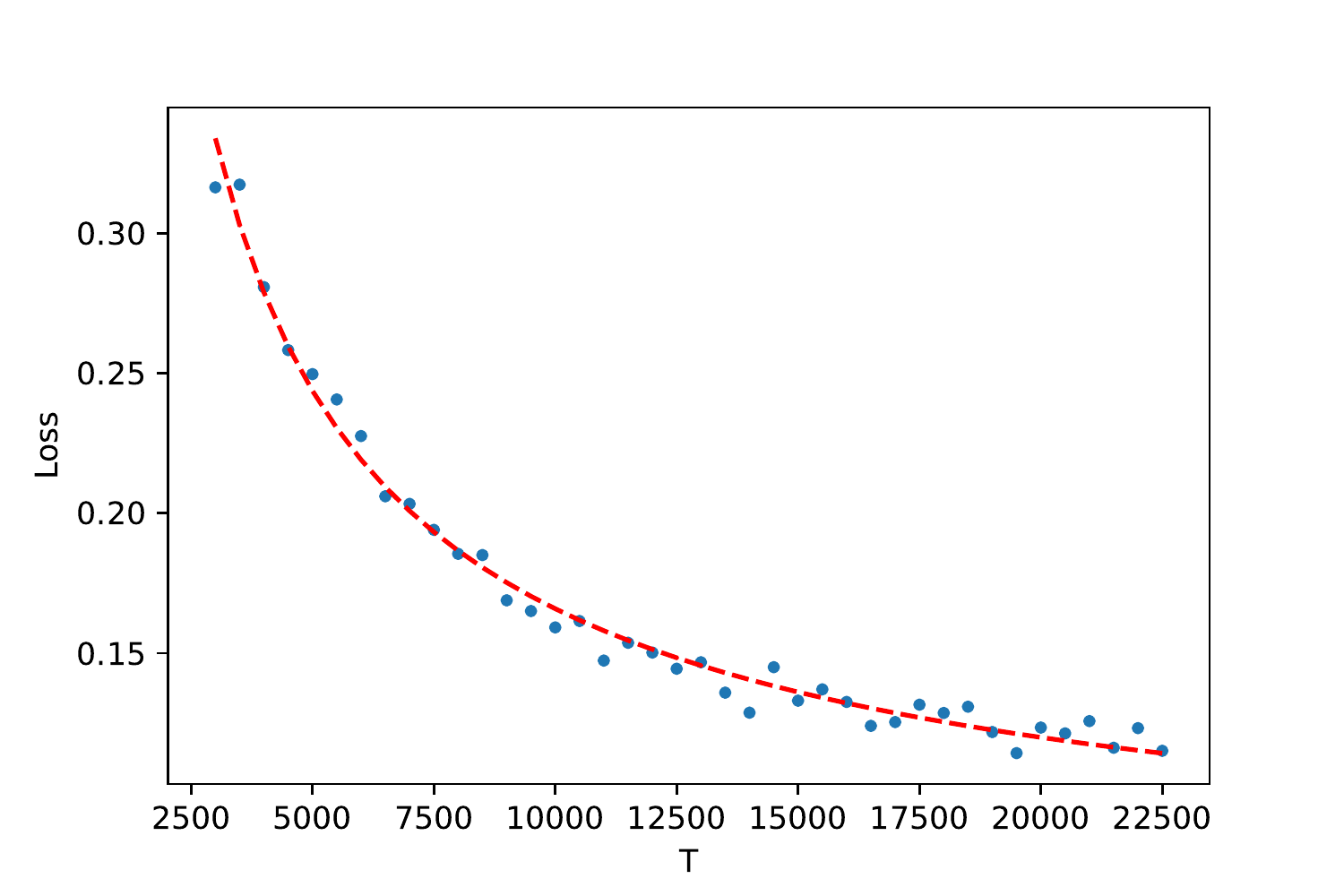}
    \caption{\textbf{MNIST Trial}}
  \end{subfigure}
  \hfill
  \begin{subfigure}[t]{.5\textwidth}
    \centering
    \includegraphics[width=\linewidth]{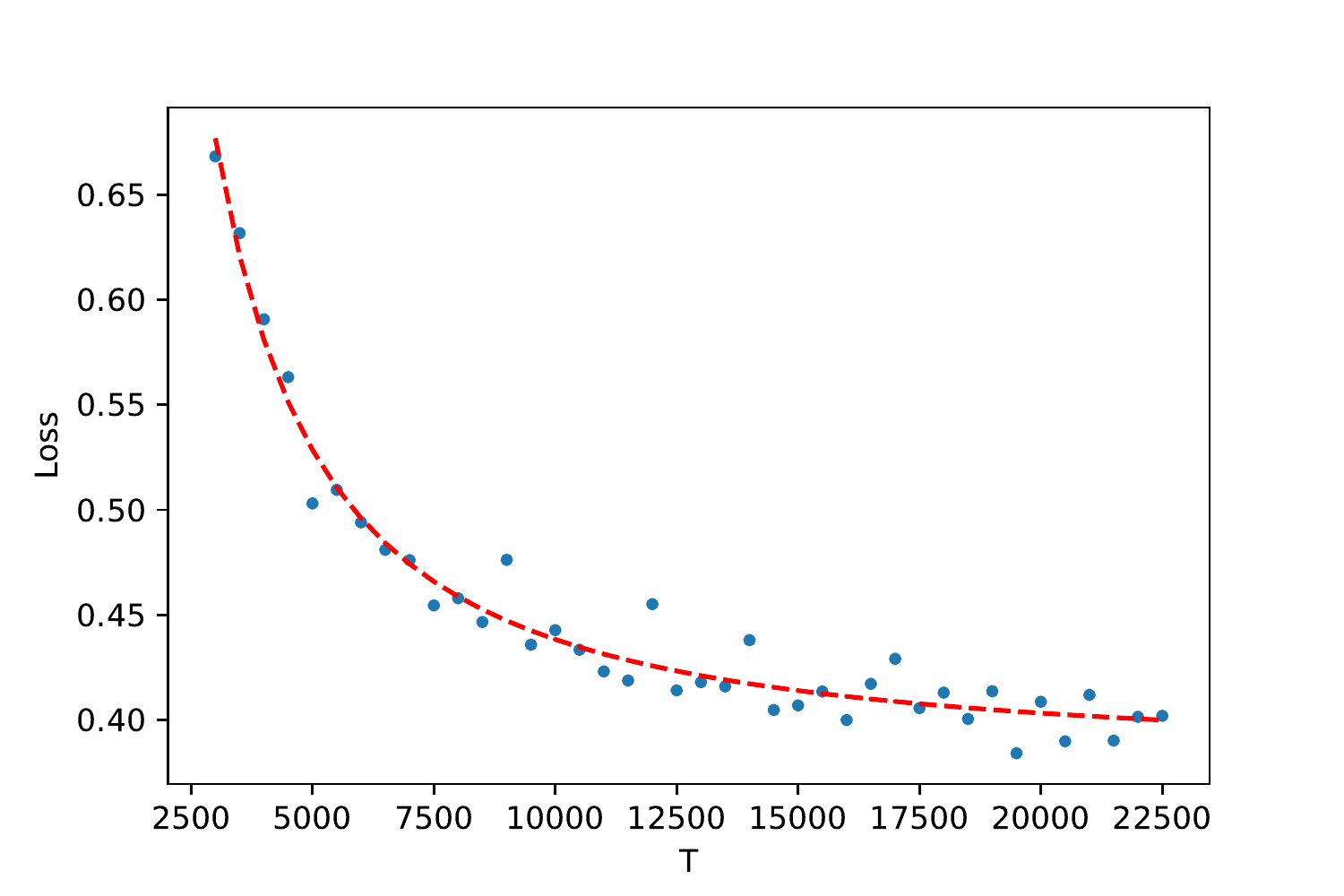}
    \caption{\textbf{Fashion-MNIST Trial}}
  \end{subfigure}
  \caption{Categorical cross-entropy loss as a function of $T$ in a simple feedforward neural network with varying amounts of trial data. Here, we display the results for a single complete run in the case of the MNIST and Fashion-MNIST data sets. In nearly all examples, we observe a highly accurate fit to a power-law behavior, with respective power-laws $103 T^{-0.74} + 0.05$ ( $R^2$ of $0.98$) and $16033 T^{-1.36} + 0.38$ ($R^2$ of $0.96$) for the MNIST and Fashion-MNIST and examples. See also table \ref{tab:NET}. We collect the central values and variance of the decay law parameters for the different data sets in table \ref{tab:AGGRO}.}
  \label{FeedForwardExperiments}
\end{figure}

The first curve is with 3000 initial samples used as training data and then incremented in steps of 500. The fit to a power-law has an $R^2$ value of 0.98, showing a very strong fit to the data with a power-law behavior $\sim 1 / T^{0.74}$. We then repeated the experiment with the Fashion-MNIST data set, which had an $R^2$ fit to power-law of 0.96 with power-law behavior $ \sim 1 / T^{1.36}$. See figure \ref{FeedForwardExperiments} for the plots of the loss function and the fitting curves, and table \ref{tab:NET} for a summary of the fitting functions for these particular examples. Table \ref{tab:AGGRO} also reports the mean and $1 \sigma$ uncertainties for the power law fitting parameters.

\begin{table}[h!]
\centering
\begin{tabular}
[c]{|l|l|l|l|l|}\hline
Dataset & Network & Function Type & Loss$(T)$ & $R^{2}$\\\hline
MNIST & FF & Power Law & $103T^{-0.74}+0.05$ & $0.98$\\\hline
Fashion-MNIST & FF & Power Law & $16033T^{-1.36}+0.41$ & $0.96$\\\hline
MNIST & CNN & Power Law & $241T^{-1.03}+0.03$ & $0.99$\\\hline
CIFAR10 & CNN & Exponential & $4.1e^{-0.000113T}+0.60$ & $0.96$\\\hline
\end{tabular}
\caption{Fitting functions categorical cross-entropy loss as a function of $T$ for the example trial runs displayed in figures \ref{FeedForwardExperiments} and \ref{CNNExperiments} for various data sets and neural network architectures (FF refers to feedforward and CNN refers to convolutional neural network). In most cases, we observe a rather good fit to a power-law behavior when the accuracy of inference is also high. For situations where there is a degraded performance as in the CIFAR10 data set, we instead observe a better fit to an exponential decay function. Note also that in some cases, we obtain a power-law with exponent above or below $-1$. Including hidden variables in the Bayesian flow equations can accommodate both phenomena. Comparing over multiple trial runs, we observe some variance in individual fits. We collect the central values and variance of the decay law parameters for the different data sets in table \ref{tab:AGGRO}.}
\label{tab:NET}
\end{table}

\begin{table}[h!]
\centering
\begin{tabular}
[c]{|c|c|c|c|}\hline
Dataset & Network & Loss($T)$ &  $b$ \\\hline
MNIST & FF & $aT^{-b}+c$  & $0.74\pm0.06$\\\hline
Fashion-MNIST & FF & $aT^{-b}+c$ &  $1.32\pm0.12$\\\hline
MNIST & CNN & $aT^{-b}+c$ & $1.01\pm0.06$\\\hline
CIFAR10 & CNN & $ae^{-bT}+c$ & $1.6 \times 10^{-4}\pm2.4 \times 10^{-5}$\\\hline
\end{tabular}
\caption{Central values of the fitting parameters averaged over $10$ different trials. Uncertainties are quoted at the $1 \sigma$ level. For the MNIST and Fashion-MNIST data sets, these fit well to power-law behavior of the form $a T^{-b} + c$. For the CIFAR10 where the overall accuracy was lower, we instead find a better fit to an exponential decay law $a e^{-bT} + c$. While there is some variance in the overall value of these fitting parameters, each individual trial fits well to the expectations of the dynamical Bayesian evolution equations. The experiments thus reveal the sensitivity to initial conditions in the training of the neural networks. }
\label{tab:AGGRO}
\end{table}

\newpage

\subsubsection{Convolutional Neural Networks}
We also performed a similar set of experiments using the convolutional neural networks as described above. We again repeated the experiments 10 times so as to take into account the stochastic nature of the training process and take mean values. We took the initial data size to be 500 and increment size 500 as before.

In the case of the MNIST data set, we find the mean power-law fit has $R^2=0.99$ and mean decay coefficient 1.01. (The root of the variance of the decay constant was 0.06).  This network had a very low final loss 0.99 and captured well the properties of the full data set. It is interesting that when this happened, the exponent of the power-law approached the value for the Cram\'{e}r-Rao bounded flow. Figure \ref{CNNExperiments}a displays one such trial. Averaging over all the trials, we also determined the exponent for the power-law decay, the results are displayed in table \ref{tab:AGGRO}.

Finally, for the CIFAR10 data set with the three layer convolutional network, we took an initial data size of 500 and increment size of 500. We repeated the experiment 10 times, and in each trial we performed a best fit to the loss function, and in general we observed the data was better fit by an exponential rather than a power-law. In figure \ref{CNNExperiments}b we present the data from one such trial, where the power-law fit (green curve) gave an $R^2$ of 0.92, while the exponential fit (red curve) gave an $R^2$ of $0.96$. Averaging over all the trials, we also determined the decay constant for the exponential fit, the results are displayed in table \ref{tab:AGGRO}. Note that although it is better fit by an exponential decay, the actual decay constant is quite small.

\section{Conclusions and Discussion} \label{sec:CONC}

In this note we have presented an interpretation of Bayesian updating in terms of a dynamical
system. In a given model of the world, each new piece of evidence provides us with an improved understanding
of the underlying system, thus generating an effective flow in the space of parameters which is saturated by a simple $1/T$ power-law, the analog of a ``unitarity bound'' in conformal field theory. This can be exceeded when additional information flows in via hidden variables. We have shown how this works in practice both in an analytic treatment of Gaussian distributions and Gaussian Random Processes, and have also performed a number of numerical experiments, including inference on the value of the coupling constants in the 1D Ising Model, and in training of neural networks. We find it remarkable that simple Bayesian considerations accurately capture the asymptotic behavior of so many phenomena.

The appearance of a $1 / T^{b}$ power-law scaling for learning in neural networks is of course quite suggestive. In the context of
statistical field theory, the onset of such a scaling law behavior is usually a clear indication of a phase transition. We have also seen that inference in the presence of hidden variables provides a simple qualitative explanation for some of this behavior. It would be very interesting
to develop a more fundamental explanation.

A unifying thread of this work has centered on giving a physical interpretation of Bayesian updating. This equation shares a number of common features with the related question of renormalization
group flow in a quantum field theory.\footnote{The notion of ``renormalization'' has been discussed in \cite{Halverson:2020trp, Halverson:2021aot, Roberts:2021fes}, though we should point out that in a quantum field theory, the utility of organizing by scale has a great deal to do with the fact that there is a clear notion of locality, something which is definitely \textit{not} present in many inference problems! For additional discussion on connections between statistical / quantum field theory and machine learning, see, e.g., \cite{Bachtis:2021xoh,Bachtis:2021pou,Aarts:2021rfa,Bachtis:2021cxh,Bachtis:2021eww}.} But whereas renormalization is usually interpreted as a flow from the ultraviolet to the infrared wherein we \textit{lose} information about microscopic physics, the Bayesian updating procedure does precisely the opposite: we are \textit{gaining} information as we evolve along a flow. We have also seen that new evidence in Bayesian updating can either perturb a trajectory, or not impact it very much, and this again parallels similar notions of relevant and irrelevant perturbations. We have also taken some preliminary steps in developing a path integral interpretation of Bayesian flows in Appendix \ref{app:PATH}. This in turn suggests that there should be a direct analog of Polchinski's exact renormalization group equation which would be exciting to develop.

One of the original motivations of this work was to better understand the sense in which the structure of quantum gravity might emerge from
an observer performing local measurements in their immediate vicinity (see, e.g., \cite{Heckman:2013kza, Balasubramanian:2014bfa, Hashimoto:2018bnb, Hashimoto:2019bih, Gal:2020dyc, Balasubramanian:2020lux} for related discussions). From this perspective, each new piece of data corresponds to this local observer making larger excursions in the spacetime, as well as the parameters of the theory. This is particularly well-motivated in the specific context of the AdS/CFT correspondence \cite{Maldacena:1997re}, where the radial direction of the bulk anti-de Sitter space serves as a renormalization scale in the CFT with a cutoff. Given that we have a flow equation, and that it shares many formal similarities to an RG equation, this suggests a natural starting point for directly visualizing radial evolution in terms of such an inference procedure.

At a more practical level, it would also be interesting to test how well an observer can infer such ``spacetime locality''. Along these lines, there is a natural class of numerical experiments involving a mild generalization of our Ising model analysis in which we continue to draw from the same Ising model with only nearest neighbor interactions, but in which the model involves additional contributions coupling neighbors which might be very far away.


\section*{Acknowledgments}

We thank J.G Bernstein, R. Fowler and R.A. Yang for helpful discussions
and many of the members of the ``Physics meets ML'' group.
DSB thanks Pierre Andurand for his generous donation supporting this work.
The work of JJH is supported in part by the DOE (HEP) Award
DE-SC0013528, and a generous donation by P. Kumar, as well as a generous donation by R.A. Yang and Google.

\newpage

\appendix

\section{Interpreting Dynamical Bayesian Updating} \label{app:PATH}

In the main text of our paper we implemented an approach to dynamical Bayesian Inference in which the posterior distribution is probed by observing the scaling of its various centralized moments as a function of update ``time". In this appendix we would like to draw attention to an alternative strategy for studying Dynamical Bayesian inference in which one solves the flow equation for the complete posterior, (\ref{Posterior Solution}), directly. As was the case in the main text, we will find it more natural to consider our update in terms of a ``time" parameter $T = N\tau$. One can think of $T$ as corresponding to the number of data point utilized in the Bayesian Inference model up to a given iteration. In these terms we can write the $T$-dependent posterior distribution which solves the flow equation as:
\begin{equation} \label{T dependent Posterior}
    \pi(\theta;T) = \exp(-T D_{KL}(\alpha_* \parallel \theta)) \exp \left(\int_{0}^T dT' D_{KL}(\alpha_* \parallel \alpha(T')) \right)
\end{equation}
We will see that the structure of this solutions calls to mind many of the common approaches utilized in the analysis of physical systems, especially statistical ensembles.

To begin, observe that $\pi(\theta;T)$ is a \emph{normalized} probability density function for each value of $T$:
\begin{equation}
    1 = \int d\theta \pi(\theta;T) \;\; \forall \; T
\end{equation}
Performing the integration explicitly, we notice that only the first factor in (\ref{T dependent Posterior}) depends on $\theta$. Thus, we find:
\begin{equation} \label{Normalization Condition}
    1 = \exp(\int_0^T dT' \, D_{KL}(\alpha_* \parallel \alpha(T'))) \int d\theta \exp(-T D_{KL}(\alpha_* \parallel \theta))
\end{equation}
It is natural to define the integral appearing in (\ref{Normalization Condition}) as the Partition Function of an unnormalized density:
\begin{equation}
    \mathcal{Z}(T) := \int d\theta e^{-T D_{KL}(\alpha_* \parallel \theta)}
\end{equation}
This gives the Dynamical Bayesian Posterior the complexion of a Bolzmann weight with ``energy" $D_{KL}(\alpha_* \parallel \theta)$. It also suggests that we should regard $T$ as an inverse temperature, or imaginary time parameter as is typical in statistical field theory contexts.

Referring back to (\ref{Normalization Condition}), we conclude that the role of the $\theta$ independent term in the posterior density (\ref{T dependent Posterior}) is explicitly to maintain the normalization of the posterior density at all $T$. Indeed, we can write:
\begin{equation} \label{Partition Function}
    \mathcal{Z}(T) = \exp \left( -\int_0^T dT' \, D_{KL}(\alpha_* \parallel \alpha(T')) \right)
\end{equation}
Or, equivalently:
\begin{equation}
    -\ln(\mathcal{Z}(T)) = \int_0^T dT' \, D_{KL}(\alpha_* \parallel \alpha(T'))
\end{equation}
This equation relates the KL-Divergence of the $T$-dependent parameter estimate $\alpha(T)$ with the cumulant generating functional of the posterior distribution. Taking the first derivative of this equation with respect to $T$ we find:
\begin{equation}
    D_{KL}(\alpha_* \parallel \alpha(T)) = \langle D_{KL}(\alpha_* \parallel \theta) \rangle_{\pi(\theta;T)}
\end{equation}
Which is precisely equation (\ref{Expectation of KL})! More generally, notice that:
\begin{equation}
    -\left( \frac{d}{dT} \right)^n \ln(\mathcal{Z}(T)) = (-1)^{n+1} \mathcal{C}^n_{\pi(\theta;T)} \left( D_{KL}(\alpha_* \parallel \theta) \right)
\end{equation}
Where here $\mathcal{C}^n_{\pi(\theta;T)}(Q(\theta))$ denotes the $n^{th}$ cumulant of $Q(\theta)$ with respect to the time $T$ posterior distribution, $\pi(\theta;T)$. We therefore obtain the expression:
\begin{equation} \label{Cumulant Equations}
    \left(\frac{d}{dT}\right)^{n-1} D_{KL}(\alpha_* \parallel \alpha(T)) = (-1)^{n+1} \mathcal{C}^n_{\pi(\theta;T)}\left(D_{KL}(\alpha_* \parallel \theta)\right)
\end{equation}
One may interpret this equation as saying that all of the relevant connected correlation functions associated with the statistical inference are encoded in the path $\alpha(T)$. Once $\alpha(T)$ is known these cumulants can extracted through equation (\ref{Cumulant Equations}).



\bibliographystyle{utphys}
\bibliography{BayesFlow}

\providecommand{\href}[2]{#2}\begingroup\raggedright\begin{thebibliography}{10}

\bibitem{Bayes:1764vd}
T.~Bayes, Rev., ``{An essay toward solving a problem in the doctrine of
  chances},'' \href{http://dx.doi.org/10.1098/rstl.1763.0053}{{\em Phil. Trans.
  Roy. Soc. Lond.} {\bfseries 53} (1764) 370--418}.

\bibitem{Gell-Mann:1954yli}
M.~Gell-Mann and F.~E. Low, ``{Quantum electrodynamics at small distances},''
  \href{http://dx.doi.org/10.1103/PhysRev.95.1300}{{\em Phys. Rev.} {\bfseries
  95} (1954) 1300--1312}.

\bibitem{Kadanoff:1966wm}
L.~P. Kadanoff, ``{Scaling laws for Ising models near $T_c$},''
  \href{http://dx.doi.org/10.1103/PhysicsPhysiqueFizika.2.263}{{\em Physics
  Physique Fizika} {\bfseries 2} (1966) 263--272}.

\bibitem{Wilson:1971bg}
K.~G. Wilson, ``{Renormalization group and critical phenomena. 1.
  Renormalization group and the Kadanoff scaling picture},''
  \href{http://dx.doi.org/10.1103/PhysRevB.4.3174}{{\em Phys. Rev. B}
  {\bfseries 4} (1971) 3174--3183}.

\bibitem{Wilson:1971dh}
K.~G. Wilson, ``{Renormalization group and critical phenomena. 2. Phase space
  cell analysis of critical behavior},''
  \href{http://dx.doi.org/10.1103/PhysRevB.4.3184}{{\em Phys. Rev. B}
  {\bfseries 4} (1971) 3184--3205}.

\bibitem{Polchinski:1983gv}
J.~Polchinski, ``{Renormalization and Effective Lagrangians},''
  \href{http://dx.doi.org/10.1016/0550-3213(84)90287-6}{{\em Nucl. Phys. B}
  {\bfseries 231} (1984) 269--295}.

\bibitem{Balasubramanian:1996bn}
V.~Balasubramanian, ``{Statistical inference, occam's razor and statistical
  mechanics on the space of probability distributions},''
  \href{http://arxiv.org/abs/cond-mat/9601030}{{\ttfamily
  arXiv:cond-mat/9601030}}.

\bibitem{KullbackLeibler}
S.~Kullback and R.~A. Leibler, ``{On Information and Sufficiency},'' {\em The
  Annals of Mathematical Statistics} {\bfseries 22} no.~1, (1951) 79 -- 86.

\bibitem{Rao}
C.~R. Rao, ``Information and accuracy attainable in the estimation of
  statistical parameters",'' {\em Bulletin of the Calcutta Mathematical
  Society} {\bfseries 37} (1945) 81--91.

\bibitem{Amari}
S.~Amari, {\em {Differential-Geometrical Methods in Statistics}}.
\newblock Lecture Notes in Statistics. Springer-Verlag, 1985.

\bibitem{Bialek:1996kd}
W.~Bialek, C.~G. Callan, Jr., and S.~P. Strong, ``{Field theories for learning
  probability distributions},''
  \href{http://dx.doi.org/10.1103/PhysRevLett.77.4693}{{\em Phys. Rev. Lett.}
  {\bfseries 77} (1996) 4693--4697},
  \href{http://arxiv.org/abs/cond-mat/9607180}{{\ttfamily
  arXiv:cond-mat/9607180}}.

\bibitem{Blau:2001gj}
M.~Blau, K.~S. Narain, and G.~Thompson, ``{Instantons, the information metric,
  and the AdS / CFT correspondence},''
  \href{http://arxiv.org/abs/hep-th/0108122}{{\ttfamily arXiv:hep-th/0108122}}.

\bibitem{Miyamoto:2012qv}
U.~Miyamoto and S.~Yahikozawa, ``{Information metric from a linear sigma
  model},'' \href{http://dx.doi.org/10.1103/PhysRevE.85.051133}{{\em Phys. Rev.
  E} {\bfseries 85} (2012) 051133},
  \href{http://arxiv.org/abs/1205.3211}{{\ttfamily arXiv:1205.3211 [math-ph]}}.

\bibitem{Heckman:2013kza}
J.~J. Heckman, ``{Statistical Inference and String Theory},''
  \href{http://dx.doi.org/10.1142/S0217751X15501602}{{\em Int. J. Mod. Phys. A}
  {\bfseries 30} no.~26, (2015) 1550160},
  \href{http://arxiv.org/abs/1305.3621}{{\ttfamily arXiv:1305.3621 [hep-th]}}.

\bibitem{Heckman:2016wte}
J.~J. Heckman, J.~G. Bernstein, and B.~Vigoda, ``{MCMC with Strings and Branes:
  The Suburban Algorithm (Extended Version)},''
  \href{http://dx.doi.org/10.1142/S0217751X17501330}{{\em Int. J. Mod. Phys. A}
  {\bfseries 32} no.~22, (2017) 1750133},
  \href{http://arxiv.org/abs/1605.05334}{{\ttfamily arXiv:1605.05334
  [physics.comp-ph]}}.

\bibitem{Heckman:2016jud}
J.~J. Heckman, J.~G. Bernstein, and B.~Vigoda, ``{MCMC with Strings and Branes:
  The Suburban Algorithm},'' \href{http://arxiv.org/abs/1605.06122}{{\ttfamily
  arXiv:1605.06122 [stat.CO]}}.

\bibitem{Clingman:2015lxa}
T.~Clingman, J.~Murugan, and J.~P. Shock, ``{Probability Density Functions from
  the Fisher Information Metric},''
  \href{http://arxiv.org/abs/1504.03184}{{\ttfamily arXiv:1504.03184 [cs.IT]}}.

\bibitem{Malek:2015hea}
E.~Malek, J.~Murugan, and J.~P. Shock, ``{The Information Metric on the moduli
  space of instantons with global symmetries},''
  \href{http://dx.doi.org/10.1016/j.physletb.2015.12.044}{{\em Phys. Lett. B}
  {\bfseries 753} (2016) 660--663},
  \href{http://arxiv.org/abs/1507.08894}{{\ttfamily arXiv:1507.08894
  [hep-th]}}.

\bibitem{Dimov:2020fzi}
H.~Dimov, I.~N. Iliev, M.~Radomirov, R.~C. Rashkov, and T.~Vetsov,
  ``{Holographic Fisher information metric in Schr\"odinger spacetime},''
  \href{http://dx.doi.org/10.1140/epjp/s13360-021-02109-0}{{\em Eur. Phys. J.
  Plus} {\bfseries 136} no.~11, (2021) 1128},
  \href{http://arxiv.org/abs/2009.01123}{{\ttfamily arXiv:2009.01123
  [hep-th]}}.

\bibitem{Erdmenger:2020vmo}
J.~Erdmenger, K.~T. Grosvenor, and R.~Jefferson, ``{Information geometry in
  quantum field theory: lessons from simple examples},''
  \href{http://dx.doi.org/10.21468/SciPostPhys.8.5.073}{{\em SciPost Phys.}
  {\bfseries 8} no.~5, (2020) 073},
  \href{http://arxiv.org/abs/2001.02683}{{\ttfamily arXiv:2001.02683
  [hep-th]}}.

\bibitem{Tsuchiya:2021hvg}
A.~Tsuchiya and K.~Yamashiro, ``{A geometrical representation of the quantum
  information metric in the gauge/gravity correspondence},''
  \href{http://dx.doi.org/10.1016/j.physletb.2021.136830}{{\em Phys. Lett. B}
  {\bfseries 824} (2022) 136830},
  \href{http://arxiv.org/abs/2110.13429}{{\ttfamily arXiv:2110.13429
  [hep-th]}}.

\bibitem{Fowler:2021oje}
R.~Fowler and J.~J. Heckman, ``{Misanthropic Entropy and Renormalization as a
  Communication Channel},'' \href{http://arxiv.org/abs/2108.02772}{{\ttfamily
  arXiv:2108.02772 [hep-th]}}.

\bibitem{Neal}
R.~Neal, {\em {Bayesian Learning for Neural Networks}}.
\newblock Lecture Notes in Statistics. Springer-Verlag, 1996.

\bibitem{Mehta:2018dln}
P.~Mehta, M.~Bukov, C.-H. Wang, A.~G.~R. Day, C.~Richardson, C.~K. Fisher, and
  D.~J. Schwab, ``{A high-bias, low-variance introduction to Machine Learning
  for physicists},''
  \href{http://dx.doi.org/10.1016/j.physrep.2019.03.001}{{\em Phys. Rept.}
  {\bfseries 810} (2019) 1--124},
  \href{http://arxiv.org/abs/1803.08823}{{\ttfamily arXiv:1803.08823
  [physics.comp-ph]}}.

\bibitem{Mack:1975je}
G.~Mack, ``{All unitary ray representations of the conformal group SU(2,2) with
  positive energy},'' \href{http://dx.doi.org/10.1007/BF01613145}{{\em Commun.
  Math. Phys.} {\bfseries 55} (1977) 1}.

\bibitem{GRP}
C.~E. Rasmussen and C.~K.~I. Williams, {\em {Gaussian Processes for Machine
  Learning}}.
\newblock The MIT Press, 2006.

\bibitem{Balasubramanian:2014bfa}
V.~Balasubramanian, J.~J. Heckman, and A.~Maloney, ``{Relative Entropy and
  Proximity of Quantum Field Theories},''
  \href{http://dx.doi.org/10.1007/JHEP05(2015)104}{{\em JHEP} {\bfseries 05}
  (2015) 104}, \href{http://arxiv.org/abs/1410.6809}{{\ttfamily arXiv:1410.6809
  [hep-th]}}.

\bibitem{Balasubramanian:2020lux}
V.~Balasubramanian, J.~J. Heckman, E.~Lipeles, and A.~P. Turner, ``{Statistical
  Coupling Constants from Hidden Sector Entanglement},''
  \href{http://dx.doi.org/10.1103/PhysRevD.103.066024}{{\em Phys. Rev. D}
  {\bfseries 103} no.~6, (2021) 066024},
  \href{http://arxiv.org/abs/2012.09182}{{\ttfamily arXiv:2012.09182
  [hep-th]}}.

\bibitem{Pathria:1996hda}
R.~K. Pathria, {\em {Statistical Mechanics}}.
\newblock Butterworth-Heinemann,
1996.
\newblock

\bibitem{Roberts:2021fes}
D.~A. Roberts, S.~Yaida, and B.~Hanin, ``{The Principles of Deep Learning
  Theory},'' \href{http://arxiv.org/abs/2106.10165}{{\ttfamily arXiv:2106.10165
  [cs.LG]}}.

\bibitem{2014arXiv1412.6980K}
D.~P. {Kingma} and J.~{Ba}, ``{Adam: A Method for Stochastic Optimization},''
  {\em arXiv e-prints} (Dec., 2014) arXiv:1412.6980,
  \href{http://arxiv.org/abs/1412.6980}{{\ttfamily arXiv:1412.6980 [cs.LG]}}.

\bibitem{Halverson:2020trp}
J.~Halverson, A.~Maiti, and K.~Stoner, ``{Neural Networks and Quantum Field
  Theory},'' \href{http://dx.doi.org/10.1088/2632-2153/abeca3}{{\em Mach.
  Learn. Sci. Tech.} {\bfseries 2} no.~3, (2021) 035002},
  \href{http://arxiv.org/abs/2008.08601}{{\ttfamily arXiv:2008.08601 [cs.LG]}}.

\bibitem{Halverson:2021aot}
J.~Halverson, ``{Building Quantum Field Theories Out of Neurons},''
  \href{http://arxiv.org/abs/2112.04527}{{\ttfamily arXiv:2112.04527
  [hep-th]}}.

\bibitem{Bachtis:2021xoh}
D.~Bachtis, G.~Aarts, and B.~Lucini, ``{Quantum field-theoretic machine
  learning},'' \href{http://dx.doi.org/10.1103/PhysRevD.103.074510}{{\em Phys.
  Rev. D} {\bfseries 103} no.~7, (2021) 074510},
  \href{http://arxiv.org/abs/2102.09449}{{\ttfamily arXiv:2102.09449
  [hep-lat]}}.

\bibitem{Bachtis:2021pou}
D.~Bachtis, G.~Aarts, and B.~Lucini, ``{Quantum field theories, Markov random
  fields and machine learning},'' in {\em {32nd IUPAP Conference on
  Computational Physics}}.
\newblock 10, 2021.
\newblock \href{http://arxiv.org/abs/2110.10928}{{\ttfamily arXiv:2110.10928
  [cs.LG]}}.

\bibitem{Aarts:2021rfa}
G.~Aarts, D.~Bachtis, and B.~Lucini, ``{Interpreting machine learning functions
  as physical observables},'' in {\em {38th International Symposium on Lattice
  Field Theory}}.
\newblock 9, 2021.
\newblock \href{http://arxiv.org/abs/2109.08497}{{\ttfamily arXiv:2109.08497
  [hep-lat]}}.

\bibitem{Bachtis:2021cxh}
D.~Bachtis, G.~Aarts, and B.~Lucini, ``{Machine learning with quantum field
  theories},'' in {\em {38th International Symposium on Lattice Field Theory}}.
\newblock 9, 2021.
\newblock \href{http://arxiv.org/abs/2109.07730}{{\ttfamily arXiv:2109.07730
  [cs.LG]}}.

\bibitem{Bachtis:2021eww}
D.~Bachtis, G.~Aarts, F.~Di~Renzo, and B.~Lucini, ``{Inverse Renormalization
  Group in Quantum Field Theory},''
  \href{http://dx.doi.org/10.1103/PhysRevLett.128.081603}{{\em Phys. Rev.
  Lett.} {\bfseries 128} no.~8, (2022) 081603},
  \href{http://arxiv.org/abs/2107.00466}{{\ttfamily arXiv:2107.00466
  [hep-lat]}}.

\bibitem{Hashimoto:2018bnb}
K.~Hashimoto, S.~Sugishita, A.~Tanaka, and A.~Tomiya, ``{Deep Learning and
  Holographic QCD},'' \href{http://dx.doi.org/10.1103/PhysRevD.98.106014}{{\em
  Phys. Rev. D} {\bfseries 98} no.~10, (2018) 106014},
  \href{http://arxiv.org/abs/1809.10536}{{\ttfamily arXiv:1809.10536
  [hep-th]}}.

\bibitem{Hashimoto:2019bih}
K.~Hashimoto, ``{AdS/CFT correspondence as a deep Boltzmann machine},''
  \href{http://dx.doi.org/10.1103/PhysRevD.99.106017}{{\em Phys. Rev. D}
  {\bfseries 99} no.~10, (2019) 106017},
  \href{http://arxiv.org/abs/1903.04951}{{\ttfamily arXiv:1903.04951
  [hep-th]}}.

\bibitem{Gal:2020dyc}
Y.~Gal, V.~Jejjala, D.~K. Mayorga~Pena, and C.~Mishra, ``{Baryons from Mesons:
  A Machine Learning Perspective},''
  \href{http://arxiv.org/abs/2003.10445}{{\ttfamily arXiv:2003.10445
  [hep-ph]}}.

\bibitem{Maldacena:1997re}
J.~M. Maldacena, ``{The Large N limit of superconformal field theories and
  supergravity},'' \href{http://dx.doi.org/10.1023/A:1026654312961}{{\em Adv.
  Theor. Math. Phys.} {\bfseries 2} (1998) 231--252},
  \href{http://arxiv.org/abs/hep-th/9711200}{{\ttfamily arXiv:hep-th/9711200}}.

\end{thebibliography}\endgroup

\end{document}